\documentclass{emulateapj}

\usepackage{graphicx}

\newcommand{\ri}{$R_\mathrm{I}$}
\newcommand{\civline}{C\,IV $\lambda$1549}
\newcommand{\ewmgii}{EW$_\mathrm{Mg\,II}$}
\newcommand{\ewciv}{EW$_\mathrm{C\,IV}$}

\newcommand{\ewoiii}{EW$_\mathrm{[O\,III]}$}
\newcommand{\alphaUV}{$\alpha^\mathrm{UV}_\nu$}

\shorttitle{Correlations of Quasar Spectra}
\shortauthors{Kimball et al.}

\begin{document}
\bibliographystyle{apj}

\title{Correlations of Quasar Optical Spectra with Radio Morphology}
\author{Amy E. Kimball\altaffilmark{1,2}}
\affil{akimball@nrao.edu}
\author{\v{Z}eljko Ivezi\'c\altaffilmark{2}}
\author{Paul J. Wiita\altaffilmark{3,4}}
\author{Donald P. Schneider\altaffilmark{5}}

\altaffiltext{1}{National Radio Astronomy Observatory, 520 Edgemont Road, Charlottesville, VA 22903-2475}
\altaffiltext{2}{Department of Astronomy, University of Washington, Box 351580,
  Seattle, WA 98195-1580}
\altaffiltext{3}{Department of Physics, The College of New Jersey, P.O.\ Box 7718,
Ewing, NJ 08628} 
\altaffiltext{4}{Department of Physics and Astronomy, Georgia State University,
  P.O.\ Box 4106,  Atlanta, GA 30302-4106} 
\altaffiltext{5}{Department of Astronomy and Astrophysics, Pennsylvania State
  University, 525 Davey Laboratory, University Park, PA 16802}

\begin{abstract}
Using the largest homogeneous quasar sample with high-quality optical spectra and robust radio morphology classifications assembled to date, we investigate relationships between radio and optical properties with unprecedented statistical power.  The sample consists of 4714 radio quasars from the Faint Images of the Radio Sky at Twenty cm (FIRST) with $S_{20}\geq2$~mJy and with spectra from the Sloan Digital Sky Survey (SDSS).  Radio morphology classes include core-only ({\it core}), core--lobe ({\it lobe}), core--jet ({\it jet}), lobe--core--lobe ({\it triple}), and {\it double-lobe}.  Electronic tables of the quasar samples, along with spectral composites for individual morphology classes, are made available.  We examine the optical colors of these subsamples and find that radio quasars with core emission unresolved by FIRST (on $\sim5\arcsec$ scale) have a redder color distribution than radio-quiet quasars ($S_{20}\lesssim1$~mJy); other classes of radio quasars have optical color distributions similar to the radio-quiet quasars.  This analysis also suggests that optical colors of $z\lesssim2.7$ SDSS quasars are not strongly ($<0.1$~mag) biased blue.

We show that the radio core-to-lobe flux density ratio ($R$) and the radio-to-optical ($i$-band) ratio of the quasar core (\ri) are correlated, which supports the hypothesis that both parameters are indicative of line-of-sight orientation.  We investigate spectral line equivalent widths as a function of $R$ and \ri, including the [O\,III] narrow line doublet and the C\,IV $\lambda1549$ and Mg\,II $\lambda2799$ broad lines.  We find that the rest equivalent widths (EWs) of the broad lines correlate positively with \ri\ at the 4$\sigma$--8$\sigma$ level.  However, we find no strong dependence of EW on $R$, in contrast to previously published results.  A possible interpretation of these results is that EWs of quasar emission lines increase as the line-of-sight angle to the radio-jet axis decreases.  These results are in stark contrast to commonly accepted orientation-based theories, which suggest that continuum emission should increase as the angle to the radio-jet axis decreases, resulting in smaller EW of emission lines (assumed isotropic).  Finally, we observe the Baldwin effect in our sample, and find that it does not depend strongly on quasar radio morphology.  \end{abstract}


\section{INTRODUCTION}
\label{sec:intro}

The unification paradigm for radio-loud active galactic nuclei (AGN) suggests that many classes of observationally-distinct sources appear dissimilar because of orientation effects \citep[e.g.][]{urryPadovani95,jackson99}.  Because we cannot (yet) visit an AGN or observe it from different angles, we often take a statistical approach to this problem, piecing together observations of many different sources to investigate this theory.  Intrinsic source properties, such as age, size, and luminosity, must also play a role in the standard AGN picture, complicating investigations of orientation measures.  The ``holy grail" of evidence for the unification paradigm would be an observational parameter that is understood to correlate directly with orientation angle.  Owing to its variety and complexity, quasar radio morphology is often considered to hold the key to unlocking the orientation mystery.

Statistical studies of radio emission from extragalactic sources have recently entered a new era, thanks to the availability of large sky-area high-resolution radio continuum surveys that are sensitive to mJy flux density levels, such as the Faint Images of the Radio Sky at Twenty cm \citep[FIRST;][]{first} survey.  In this paper, we present a sample of 4714 radio quasars, spectroscopically-confirmed by the Sloan Digital Sky Survey \citep[SDSS;][]{york} and with robust visual classifications of radio morphology from FIRST images.  This is the largest sample of visually morphologically-classified radio quasars to date.  We use the radio quasar sample to investigate the relationship between radio morphology and optical spectral parameters in the context of orientation theories.

In standard unification theory, the basic picture of a radio-loud quasar is that of a galaxy with an accreting central black hole (an AGN) surrounded by a dust torus in the plane of the accretion flow, and with powerful relativistic jets directed outward along the rotation axis of the black hole.  The anisotropic emission leading to orientation effects is thought to have two principle causes: obscuration of the central object and immediate environs when viewed through the dust torus, and relativistic boosting of the jet emission when the line of sight is close to the jet axis.  These effects have visible consequences in both the radio and the optical regimes.  Simple models of AGN core emission motivate descriptions of how radio properties, such as morphology and spectral slope, depend on orientation.  Theories of other orientation indicators, such as spectral lines, have been empirically motivated by observed correlations with the aforementioned radio parameters.

Determining the orientation angle to any individual AGN is extremely difficult, if not impossible.  \citet{ghiselliniEtal93} discuss a possible method, which requires high-resolution radio observations of the relativistic jets.  The viewing angle is related to the apparent jet velocity and the relativistic Doppler factor, $\delta$.  An estimate of $\delta$ is based on the ``classical'' condition that the synchrotron self-Compton flux\footnote{``Synchrotron self-Compton" refers to inverse Compton scattering of synchrotron photons off of the electron population that is producing the synchrotron radiation.} from the jet cannot exceed the observed flux at X-ray frequencies.  The self-Compton flux depends on $\delta$, on the angular size of the boosted core component, and on the Compton self-absorption frequency.  The latter two properties can be determined from multi-frequency observations with high spatial resolution using Very Long Baseline Interferometry (VLBI); the necessary observations, however, are prohibitively expensive in terms of time and resources.  Furthermore, Doppler factors (and therefore derived inclination angles) measured on parsec scales can differ dramatically from those derived on kiloparsec scales owing to even modest changes in jet acceleration or direction \citep[e.g.,][]{harrisKrawczynski06}.

A common approach to orientation studies is to use large data samples and {\it statistical} indicators of orientation.  One such parameter that has been employed to great success is the radio core-to-lobe ratio, $R$.  The extended emission from AGN radio lobes is thought to be isotropic, while emission from the relativistic jets is Doppler boosted.  The parameter $R$ should therefore depend strongly on orientation, with high $R$ indicating a small angle to the jet axis \citep{orrBrowne82,kapahiSaikia82,morisawaTakahara87}.  Many other studies, considered both individually and collectively, support the use of $R$ as a statistical indicator of orientation.  \citet{antonucciUlvestad85} found that $R$ correlates with optical polarization, variability of the core, and one-sided radio morphology.  \citet{zirbelBaum95} found that $R$ correlates with total radio power in double-lobed radio sources, but with significant scatter (2 orders of magnitude).  \citet{morgantiEtal97} found higher values of $R$ in sources thought to be close to face-on as determined by radio morphology.  \citet{chiabergeEtal99} and \citet{hardcastleWorrall00} found that $R$ correlates with core radio luminosity, while \citet{kharbShastri04} found a correlation with core optical luminosity.

The scatter in these correlations with $R$ indicate that it is not a direct measure of orientation for any {\it individual} source, and that some other factor or factors must also contribute.  Age is a likely factor, as quasar radio lobes are expected to expand and dim over time \citep[e.g.][]{brw,barai07}.  Environment may also affect extended radio emission, as demonstrated by the existence of HYbrid MOrphology Radio Sources (HYMORS) \citep{gkw00}. HYMORS are radio-loud quasars having a different \citet{FR} (FR) morphology on either side of the core.  FR~I sources have two visible jets that have lost their bulk relativistic velocities on large ($>10$ kpc) scales \citep[e.g.,][]{bicknell95} and that often terminate in diffuse lobes; by definition, the majority of their emission arises from the inner half of their total extent.  FR~II sources are edge-brightened as their emission predominantly comes from their lobes; they are thought to have jets that remain relativistic out to very large distances and terminate in hot-spots \citep[e.g.,][]{gkw06}.  FR~I morphology is typical in sources with radio power $P<10^{24-25}\,\mathrm{W\,Hz}^{-1}$ while FR~II sources typically have $P>10^{24-25}\,\mathrm{W\,Hz}^{-1}$; the actual transition luminosity rises with optical luminosity of the host galaxy \citep{ledlowOwen96,gkw01b}.  Environmental asymmetries could lead to different jet interactions on either side of the central source, even if the oppositely-directed jets are similar in power and velocity \citep{gkw00}.  The majority of quasars in the sample we present are, unfortunately, too distant for accurate determination of FR class from their radio images.
 
It has been suggested that the radio-to-optical ratio of the quasar core, $R_\mathrm{V}$, is a better statistical measure of core-boosting, and therefore of orientation, than $R$ \citep{willsBrotherton95, kharbEtal10}.  $R_\mathrm{V}$ measures core-boosting by using the core's optical flux to normalize the observed (boosted) core emission.  In contrast, $R$ normalizes core-boosting using the extended radio emission; $R$ is therefore sensitive to the jet's environmental interactions far from the nucleus, while $R_\mathrm{V}$ is not.  It is important to note that using $R_V$ as a statistical parameter assumes that the optical core emission originates in the accretion disk.  For face-on sources such as blazars, viz., optically-violent variables and BL Lac objects, the beamed synchrotron emission dominates the optical spectrum.  Such sources could therefore contaminate our analysis.  However, at such extreme angles, the beamed synchrotron should overpower the line emission entirely; the fact that SDSS quasars are identified as having one or more very broad lines suggests that the optical emission is dominated by the accretion disk rather than a strongly beamed continuum.  For this paper, we measure the radio-to-optical ratio of the quasar core using the $i$ band ($\sim7481$\AA).  We will refer to this ratio as \ri\ rather than $R_\mathrm{V}$. 

Orientation is thought to affect the optical spectrum through orientation-dependent obscuration \citep[e.g.,][and references therein]{nenkovaEtal08} and/or inclination-dependent emission from an accretion disk surrounding the central black hole of an AGN \citep[e.g.,][]{baker97}.  Evidence of these effects is typically found via optical parameters, such as emission line profiles or spectral index, that correlate with the known radio signatures of orientation: $R$ and spectral slope.  For example, \citet{bakerHunstead95} made spectral composites for sets of quasars grouped by $R$ and found that as the supposed viewing angle to the jet axis increases, the optical continuum steepens, the Balmer decrement increases, and line widths increase.  They concluded that reddening is strongest in quasars that are more lobe-dominated, in accordance with the basic unification theory.

In this paper, we revisit spectral line correlations with these two proposed orientation parameters, $R$ and \ri, using a large quasar sample with well-determined core and lobe flux densities.  The remainder of the paper is laid out as follows.  In Section~\ref{sec:data}, we discuss the data selection and compilation of the radio quasars sample.  In Section~\ref{sec:morphology}, we discuss how automated estimates of quasar morphology motivate the more detailed classification presented in this paper, and describe the radio morphology visual classification method; we also investigate the intrinsic color distribution of quasars as a function of radio morphology.  Section~\ref{sec:core-boosting} investigates how $R$ and \ri\ are correlated.  Section~\ref{sec:spectra} presents a set of quasar spectral composites based on morphology, on $R$, and on \ri.  In Section~\ref{sec:line_profiles}, spectral line properties in individual spectra are investigated quantitatively as a function of the two proposed orientation parameters.  We summarize our results in Section~\ref{sec:discussion}.  Throughout, we use a cosmology with $H_0=70\,\mathrm{km\,s}^{-1}\,\mathrm{Mpc}^{-1}$, $\Omega_M=0.3$, and $\Omega_\Lambda=0.7$.

\section{DATA}
\label{sec:data}

We describe below the radio and optical sky surveys from which we draw our quasar sample.  The radio survey, FIRST, is the deepest large-area radio sky survey undertaken to date.  Performed at 20~cm, it is sensitive to the high-frequency quasar core emission; the survey contains high-resolution radio images.  The SDSS provides the largest-ever homogenous sample of spectroscopically-confirmed quasars.  FIRST was designed to cover the same region of sky as the SDSS, making this combination of sky surveys a powerful tool for statistical studies of radio quasars.

\subsection{FIRST}
\label{subsec:first}

The FIRST\footnote{{\tt http://sundog.stsci.edu}} survey \citep{first} used the NRAO\footnote{The National Radio Astronomy Observatory is a facility of the National Science Foundation operated under cooperative agreement by Associated Universities, Inc.} Very Large Array (VLA) telescope to observe the sky at 20\,cm (1.4\,GHz) with a beam size of 5\farcs4 and an rms sensitivity of about 0.15\,mJy~beam$^{-1}$.  FIRST observed $9,000 \deg^2$ at the north Galactic cap and a smaller $\sim2.5\degr$ wide strip along the Celestial Equator in the South Galactic cap, overlapping the SDSS footprint.  FIRST is 95\% complete to 2 mJy and 80\% complete to the survey limit of 1\,mJy \citep{first}.  The survey contains over 800,000 unique sources, with astrometric uncertainty of $\lesssim1\arcsec$.  From fitting a two-dimensional Gaussian to each co-added image, FIRST measured the peak flux density at 20 cm ($S_\mathrm{peak}$) and the integrated flux density at 20 cm ($S_{20}$) for each source.

\subsection{The Sloan Digital Sky Survey quasar catalog}
\label{subsec:sdss}

We draw our quasar sample from the quasar catalog presented by \citet[][hereafter S07]{dr5quasars}, which is based on the fifth data release (DR5) of the SDSS\footnote{\tt http://www.sdss.org} \citep{DR5}.  Technical details about the survey can be found in \citet{york,gunnTelescope,fukugita96,gunn98,hoggPhoto01,lupton02,scranton,ugriz,ivezic04,tucker06,padmanabhan}.  The astrometric accuracy of the SDSS is $\lesssim0.1\arcsec$ \citep{sdss_astrometry}.

A small fraction of SDSS photometric sources are selected for spectroscopic observation according to several targeting algorithms.  The quasar algorithm \citep{richards02} targets all $15<i<19.1$ (for $i$-band apparent magnitude $i$) unresolved sources within $2\arcsec$ of a FIRST radio detection.  Some quasars are targeted based on their broadband SDSS colors; others are targeted fortuitously via the algorithms for galaxies \citep{eisenstein,strauss,tiling}.

The S07 sample contains 77,429 sources from $\approx5740 \deg^2$ of sky with $i$-band absolute magnitude $M_i<-22$ and at least one emission line having FWHM larger than 1000 km s$^{-1}$, or which contain complex, broad absorption features.  Rest-frame $i$-band absolute magnitudes were calculated assuming a power-law spectral energy distribution with spectral index $\alpha_\mathrm{opt}=-0.5$.  An updated quasar catalog has recently been published \citep[][hereafter S10]{dr7quasars} from the SDSS seventh data release \citep[DR7;][]{dr7}.  In the updated DR7 catalog, 171 objects from DR5 were dropped, typically because of small changes in photometry or because their spectra do not satisfy the minimum line width as measured using a principal component analysis \citep{dr7quasars}.  We exclude these 171 objects from our analysis.  We use the redshifts, astrometry, and photometric measurements from S10.
 
\subsubsection{Spectral line measurements}
\label{subsubsec:lines}

To obtain spectral line measurements, we use the catalog of SDSS DR7 quasar properties presented by \citet{shenEtal10}.  The catalog provides emission line measurements for H$\alpha$, H$\beta$, \civline, Mg\,II $\lambda$2799, and the [O\,III] doublet at $\lambda$4959 and $\lambda$5007 (among others).  Each line was modeled using one or more Gaussian profiles, as appropriate; broad and narrow components of the emission lines were fit separately.  This method is more robust than that used by the SDSS pipeline, which fits a single Gaussian to each line profile.  The Shen et al. fits deteriorate for low-quality spectra; for the present paper, we limit our emission line analysis to spectra with median signal-to-noise ratio per pixel (69 km s$^{-1}$ SDSS pixels) $>5$.

The presence of broad absorption lines (BALs), which are evidence of quasar outflows and are at least 2000~km~s$^{-1}$ wide by definition \citep{weymann91}, can skew spectral line measurements.  \citet{gibson09_bal} visually examined each DR5 SDSS quasar spectrum and identified those with BALs near the Si~IV, C\,IV, Al~III, and Mg\,II lines.  For analysis of individual lines, we exclude quasars with a BAL associated with that wavelength.  Because the presence of a BAL can also affect the spectral shape in the ultraviolet regime, we exclude every BAL quasar from our analysis of spectral slopes in Section~\ref{subsec:color excess}; we also exclude BALs from our composite spectra in Section~\ref{subsec:composites}.

\section{Radio morphology of quasars}
\label{sec:morphology}

Radio morphology is crucial for categorizing quasars and understanding their underlying physical properties.  The apparent morphology of a radio source is determined by intrinsic parameters such as age, size, and jet power, for example, but also by orientation.  In this section, we present the classification of radio quasars by radio morphology and define the morphology classes used throughout the rest of this paper.  We then discuss the fraction of highly-reddened quasars as a function of morphology class.

\subsection{Lessons from automated morphology classification}

We begin by giving a brief overview of work by \citet[][hereafter KI08]{ki08} to classify radio sources by morphology and to describe the optical and radio properties of the different morphology classes.  Their work motivates the more detailed morphology investigations we now pursue.

KI08 compiled a large, homogeneous catalog of radio and optical sources for the purpose of exploring the multi-wavelength properties of AGNs.  A simple method was used for automatic morphology classification based on flux densities obtained at varying spatial resolution from different radio sky surveys.  They used this method to define three morphology classes: ``compact" (unresolved at 5\arcsec), ``resolved" (resolved at 5\arcsec\ but fairly concentrated), and ``complex" (having significant extended emission beyond 5\arcsec).  KI08 found that these three classes, defined entirely by their radio emission, correspond to different types of optical sources (with significant overlap among the classes).  For example, the complex and resolved classes consist of optically-identified galaxies with steep radio spectra, while flat-spectrum quasars typically fall into the compact class.  These results are consistent with the basic unification theory of radio-loud objects, which suggests that radio AGNs viewed along the jet axis appear as compact and flat-spectrum in the radio (because core-boosting dominates the radio emission) and as quasars in the optical (because the central object is unobscured).  Conversely, AGNs viewed perpendicular to the jet axis tend to be dominated by steep-spectrum extended radio emission, and appear as galaxies in the optical regime.

With the new quasar classifications presented here, we extend the analysis of KI08 by examining the dependence of quasar spectral properties on radio morphology.  Additionally, our classifications separate the ``complex" class of KI08 into finer morphology categories.

\begin{figure}
\epsscale{1.15}
\plotone{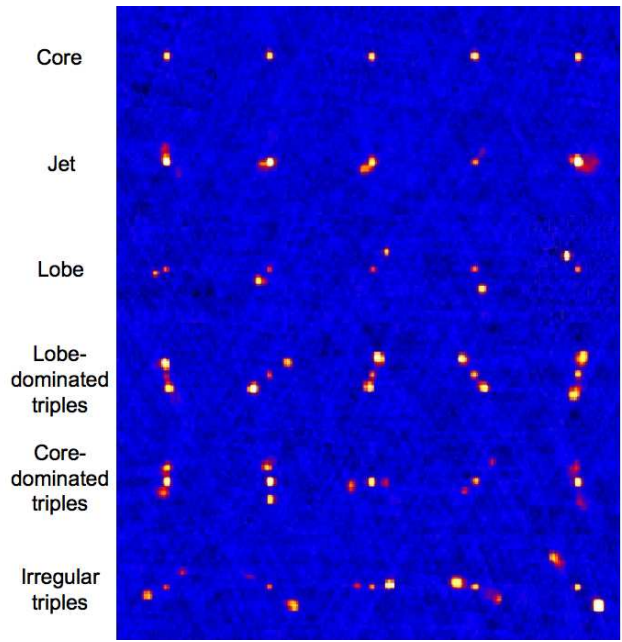}
\figcaption{\label{fig:mosaic}
Examples of the different morphology classes as labeled (see text for
definitions) in FIRST $2\arcmin\times2\arcmin$ stamps normalized to the
brightest pixel and scaled linearly.  FIRST instrumental resolution is
$\sim5\arcsec$.  The fourth and fifth rows show sources with increasing
core-to-lobe flux density ratio from left to right.}
\end{figure}

\subsection{Radio morphology classification of quasars}
\label{subsec:morphology}

\subsubsection{Sample Definition}

We defined our initial quasar sample using 77,258 quasars from S07 (the full catalog of 77,429 sources from the SDSS DR5 catalog, less the 171 sources that were dropped from the updated DR7 catalog; see \S\ref{subsec:sdss}).  We divided the S07 sample into a main sample of ``radio quasars" and a secondary sample of ``radio-quiet quasars".  The radio sample consists of 4714 quasars with a nearby FIRST source whose integrated flux density is $S_{20}\geq2$~mJy, in order to ensure a sufficiently bright sample for reliable morphology classification.  The radio-quiet sample comprises 65,253 quasars that lie in the area of FIRST coverage but do not have a FIRST counterpart.\footnote{The term ``radio-quiet" has more than one definition in the scientific literature.  It has been used in reference to a limiting radio luminosity \citep[e.g.,][]{dunlopEtal93}, as well as to a limiting radio-to-optical flux density ratio \citep[e.g.,][]{peacockEtal86}.  The definition employed here is different from both of these, as our ``radio-quiet" sample is defined by falling below the limiting radio flux density of FIRST: $\sim1$~mJy.}  We exclude from our analysis an additional 5525 quasars that lie outside the FIRST survey area.  We also exclude 1449 ``faint radio'' quasars with a faint ($S_{20}<2$~mJy) FIRST counterpart within 2\arcsec.  

The radio sample may contain a small number of false matches.  The distribution of FIRST---SDSS positional offsets is a Rayleigh distribution with $\sigma=0.2$ and a small tail out to 2\arcsec\ \citep{dr5quasars}, indicating that the number of spurious FIRST---SDSS matches within 2\arcsec\ is small.  The FIRST survey density is $\sim63\deg^{-2}$ for sources brighter than 2 mJy.  It follows that the false matching rate in the S07 catalog is $6.1\times 10^{-5}$ per source.  We therefore expect no more than 4--5 spurious radio sources in our radio quasar sample.

\subsubsection{Classification of radio images}

We classified the main sample by visual examination of $2\arcmin\times2\arcmin$ FIRST images centered on the quasar optical position.  The great majority of radio quasars are smaller than $2\arcmin$ in angular extent \citep[][hereafter dV06]{devriesW}.  For images that showed emission $\gtrsim1\arcmin$ from the optical position, we also reviewed a $4\arcmin\times4\arcmin$ field.  We categorized the images into five morphology classes: 
\begin{enumerate}
\item \noindent \emph{core}, a quasar with radio emission only at the optical position;
\item \emph{jet}, a source with radio emission from the quasar core and a radio jet;
\item \emph{lobe}, a source with radio emission from the quasar core and from a lobe;
\item \emph{triple}, a quasar with lobe--core--lobe morphology; 
\item \emph{knotty}, a single source with many emission regions or a complex emission map.
\end{enumerate}
Example images of the first four of these classes are shown in Figure~\ref{fig:mosaic}.

Two of us (AEK and PJW) examined the 4714 images, and agreed on the morphology classification of 4460 sources (95\%).  We assigned these sources a quality flag value of 1.  Author \v{Z}I examined the remaining 254 sources and acted as a tie-breaker in 175 cases; we assigned these sources a quality flag value of 2.  The remaining 84 images were sufficiently complicated that either all three examiners classified them differently, or agreed that a definite classification is not possible; these objects are listed as ``unclassified" with a quality flag value of 3.  The number of objects in each morphology class is listed in Table~\ref{table:numbers}.  The main radio quasar sample, including morphology classifications, is given in Table~\ref{table:sample}.

\setlength{\tabcolsep}{2pt}

\begin{deluxetable}{lr}
\tablewidth{2.5in}
\tablecaption{\label{table:numbers}
Quasar Classes}
\tablehead{\colhead{Sample} & \colhead{Sample size}}
\startdata
All quasars & 77,258 \\
......Excluded: outside FIRST area & 5525 \\
......Excluded: 1 mJy~$<S_{20}<2$~mJy & 1449 \\
......Radio-quiet & 65,253 \\
......Double-lobed w/o a radio core & 317 \\
\smallskip
......Main radio sample & 4714 \\
\hline
...............Core & 3433 \\
........................Unresolved & 2802 \\
\smallskip
........................Resolved & 631 \\
\smallskip
...............Jet & 183 \\
...............Lobe & 387 \\
........................Core-dominated & 233 \\
........................Lobe-dominated & 150 \\
\smallskip
........................(undetermined) & 4 \\
...............Triple & 619 \\
........................Core-dominated & 164 \\
........................Lobe-dominated & 226 \\
........................Irregular & 129 \\
........................(undetermined) & 100 \\
\smallskip
...............Knotty & 8 \\
...............unclassified & 8 \\
\enddata
\end{deluxetable}

\setlength{\tabcolsep}{6pt}

\subsubsection{Redshift Selection Effects}
\label{subsubsec:redshift effects}

The radio quasars range from 0.78 to 5.4 in redshift.  Redshift can affect the morphology classification because faint radio components drop below the FIRST detection limit at large recessional distance.  Furthermore, extended source components may be unresolved at high redshift.  Figure~\ref{fig:redshifts} shows the redshift distribution for each morphology class, and verifies that the fraction of sources with extended morphology (lobe, jet, or triple) is higher at low redshifts.

Core sources can be divided into two categories, based on whether a source is resolved or unresolved in FIRST.  From KI08, we define a dimensionless concentration parameter as the ratio of integrated flux density ($S_{20}$) to peak flux density ($S_\mathrm{peak}$) at 20~cm, according to the formula
\begin{equation}
\label{eq:theta}
\theta=\left(\frac{S_{20}}{S_\mathrm{peak}}\right)^{1/2}.
\end{equation}
``Resolved" core sources have $\log(\theta^2)>0.05$ and ``unresolved" sources have $\log(\theta^2)<0.05$.  The resolved sample likely consists of quasars with extended radio emission that are still young, and therefore small, and/or at high-redshift such that the two radio lobes appear as one component.  The unresolved sample should contain quasars with boosted radio cores, which will appear as point sources, and that do not have detectable extended emission.  However, some of the unresolved sources, like the resolved cores, may instead be distant quasars whose extended emission is not resolved on the $\sim$5\arcsec\ scale.

\subsubsection{Sub-classification of Triple and Lobe Sources}
\label{subsubsec:triples}

To measure the core-to-lobe ratio $R$, we must determine the flux densities of individual core and lobe components in the triple and lobe morphology classes.  We identified these components directly from the FIRST images, where possible.  Positive identifications were possible in 519 (383) out of 619 (387) sources with triple (lobe) morphology.  In the remaining 100 (4) images, the field is too crowded to positively identify the lobe components associated with the quasar.  For sources with multiple emission points in a lobe (162 triples; 22 lobes) we use the sum of contributing components.

We define ``lobe-dominated" sources (lobes brighter than the core), ``core-dominated" sources (lobes fainter than the core), and ``irregular" sources (one lobe brighter than the core and the other fainter; this category only applies to triple-morphology sources).  The bottom three rows of Figure~\ref{fig:mosaic} show examples of these three sub-classes among the triples.  The number in each subclass is listed in Table~\ref{table:numbers}.  The 20 cm flux densities of the cores and lobes are listed in Table~\ref{table:sample}.

\begin{deluxetable*}{rrrrcccrrr}[b]
\tablecaption{\label{table:sample}Quasar sample with radio morphology classifications}
\tablewidth{6in}
\tablehead{\colhead{SDSS name} & \multicolumn{2}{c}{R.A.\tablenotemark{a}~~(J2000)~~Dec\tablenotemark{a}} & \colhead{$z$} & \colhead{class\tablenotemark{a}} & \colhead{subclass\tablenotemark{b}} & \colhead{flag\tablenotemark{c}} & \colhead{$S_\mathrm{core}$} & \colhead{$S_\mathrm{lobe1}$\tablenotemark{d}} & \colhead{$S_\mathrm{lobe2}$\tablenotemark{d}} }
\startdata
000017.38$-$085123.7 &    0.072423 &   $-$8.856608 &   1.2491 & J & --- &        1 &      9.77 & --- & --- \\
000028.82$-$102755.7 &    0.120087 &  $-$10.465497 &   1.1518 & C &   U &        1 &      2.23 & --- & --- \\
000050.60$-$102155.9 &    0.210837 &  $-$10.365531 &   2.6404 & C &   U &        1 &     20.39 & --- & --- \\
000051.56+001202.5 &      0.214855 &      0.200707 &   3.9713 & C &   U &        1 &      2.99 & --- & --- \\
\hline
000054.96+010143.4 &      0.229009 &      1.028724 &   1.4617 & C &   U &        1 &      3.74 & --- & --- \\
000111.19$-$002011.5 &    0.296662 &   $-$0.336539 &   0.5179 & J & --- &        1 &     25.00 & --- & --- \\
000221.11+002149.3 &      0.587988 &      0.363706 &   3.0699 & C &   U &        1 &     13.88 & --- & --- \\
000222.47$-$000443.5 &    0.593646 &   $-$0.078752 &   0.8106 & C &   U &        1 &      3.89 & --- & --- \\
\hline
000442.18+000023.3 &      1.175791 &      0.006480 &   1.0068 & L &   C &        1 &      3.71 & 3.45 & --- \\
000507.05$-$101008.7 &    1.279415 &  $-$10.169089 &   1.2953 & C &   R &        1 &     94.62 & --- & --- \\
000608.04$-$010700.7 &    1.533519 &   $-$1.116869 &   0.9486 & T &   L &        1 &      4.06 & 51.14 & 29.79 \\
000622.60$-$000424.4 &    1.594198 &   $-$0.073455 &   1.0377 & C &   R &        1 &   3879.24 & --- & --- \\
\enddata
\tablecomments{This subset of the table demonstrates format and content.  The complete table is available in the electronic version of this paper.  Flux densities are given in mJy units.}
\tablenotetext{a}{SDSS position in units of decimal degrees.}
\tablenotetext{b}{C: core; J: jet; K: knotty; L: lobe; T: triple; X: unclassified; ---: no subclass.  (Section~\ref{subsec:morphology})}
\tablenotetext{c}{U: unresolved core sources; R: resolved core sources; C: core-dominated triple or lobe sources; L: lobe-dominated triple or lobe sources; I: irregular triple sources; X: undetermined subclass for triples/lobes; ---: source class has no subclass}
\tablenotetext{d}{A flag of 1 indicates a source with recognizable morphology, such that the two first-round viewers initially agreed on the category.  A flag of 2 indicates a more difficult case, where a third viewer's opinion was necessary as a tie-breaker.  A flag of 3 indicates a source with undetermined morphology.}
\tablenotetext{e}{Total observed 20 cm flux density of the extended emission (lobes) of {\it triple}- and {\it lobe}-morphology sources.  Sources with undetermined lobe components (due to crowded FIRST images) are given a default value of -99.}
\end{deluxetable*}

\subsection{Double-lobed quasars without a radio-detected core}
\label{subsec:doubles}

Not all quasars with extended radio emission have an observed radio core.  Unification theory suggests that such sources have jets aligned close to the plane of the sky.  This population is excluded from our main radio sample by the requirement of a FIRST counterpart within 2\arcsec\ of the quasar's optical position.  The fraction of omitted sources is small: \citet{lu07} estimated that 8\% of radio quasars are missed when using a 2\arcsec\ optical---radio matching radius.  These so-called {\it double-lobed} quasars are useful to test for selection bias in the SDSS spectroscopic targeting algorithm (\S\ref{subsec:color excess}).

To find a sample of double-lobed quasars, we started with our radio-quiet sample (quasars having no FIRST counterpart within 2\arcsec).  Potential lobe components were FIRST sources with no optical counterpart within 3\arcsec, and which were not associated with one of the quasars in our triple or lobe classes.  We matched the radio-quiet quasars to these FIRST sources within 120\arcsec (dV06).  For simplicity, we avoided crowded fields by limiting our focus to quasars with no more than two potential FIRST lobes within 120\arcsec.  Among the remaining candidates, we required a lobe-core-lobe opening angle of greater than 120\degr\ and a lobe-lobe flux density ratio between 0.1 and 10.  These choices of constraints were motivated by the selection algorithm for optically-faint triples (\S\ref{sec:core-boosting}; described in detail in the Appendix).  Our criteria result in a sample of 317 double-lobed quasars.

A different, more extensive method to identify double-lobed quasars even in very crowded environments is described by dV06.  For simplicity, we limit our selection to quasars in sparse environments, such that the lobes are easy to identify.  (dV06 identified optical positions of their candidates, but not the individual lobe components.)  We compare our sample to the dV06 double-lobed quasar catalog to estimate our completeness.  The dV06 sample was drawn from the SDSS Third Data Release \citep{dr3} and includes 143 quasars that are also in the S07 catalog.  Our double-lobed sample and the dV06 sample have 66 objects in common.  The remaining 251 quasars in our sample are from later SDSS data releases than the dV06 sample.  Of the 77 sources which are in dV06 but not in our sample, 12 were excluded because one of the potential FIRST lobes was associated with either an optical source or the extended emission of a triple-morphology quasar.  The remaining 63 dV06 quasars were excluded from our sample because they lie in a crowded field (more than 2 FIRST matches within 120\arcsec).  We successfully recovered 66 dV06 quasars and miss no more than 63; the lower limit on the completeness of our sample is approximately 50\%.  We note that we are biased against quasars in crowded radio sky regions, and those with lobes that were resolved into separate FIRST detections.

We use the results of \citet{lu07} as a separate estimate of the completeness of the double-lobe sample.  Lu et al. found that 8\% of radio quasars do not have a radio-core within 2\arcsec\ of the optical position.  The size of our main sample is 4714 quasars, suggesting that there are $\sim410$ corresponding FIRST---SDSS quasars that have only extended radio emission.  We found 317 double-lobed quasars, implying that our selection criteria have excluded a further $\sim93$ sources.  Based on these values, our double-lobed sample completeness approaches 77\%.  These sources are listed in Table~\ref{table:double-lobed}.

\begin{deluxetable*}{rrrrrrrrr}
\tablewidth{5in}
\tablecaption{\label{table:double-lobed}
317 Candidate double-lobed quasars without a detected radio core}
\tablehead{& & \multicolumn{2}{c}{SDSS position} & \colhead{} & \multicolumn{3}{c}{FIRST matches} \\
\colhead{SDSS name} & \multicolumn{2}{c}{R.A.~~(J2000)~~Dec} & \colhead{$z$} & \multicolumn{2}{c}{R.A.~~(J2000)~~Dec} & \colhead{$S_\mathrm{core}$ [mJy]}}
\startdata
000657.63$-$010358.8 &  1.74016 &  $-$1.06632 &  1.4358 &   1.73900 &  $-$1.06724 &    16.17 \\
                     &          &             &         &   1.74125 &  $-$1.06559 &    22.80 \\
001138.43$-$104458.2 &  2.91015 & $-$10.74952 &  1.2735 &   2.91197 & $-$10.74503 &    27.24 \\
                     &          &             &         &   2.90913 & $-$10.75229 &    60.72 \\
002909.13$-$092110.1 &  7.28805 &  $-$9.35281 &  1.5390 &   7.26567 &  $-$9.34251 &     1.88 \\
                     &          &             &         &   7.30512 &  $-$9.36770 &     9.34 \\
002948.54+004447.5   &  7.45228 &     0.74655 &  1.0078 &   7.45183 &     0.74489 &    80.55 \\
                     &          &             &         &   7.45288 &     0.75012 &    36.91 \\
003236.76$-$001446.6 &  8.15318 &  $-$0.24629 &  1.9073 &   8.14862 &  $-$0.23908 &     8.15 \\
                     &          &             &         &   8.17664 &  $-$0.26754 &     1.17 \\
\enddata
\tablecomments{The full version of this table is available in the electronic version of this paper.  Right ascension and declination are given in units of decimal degrees.}
\end{deluxetable*}

\begin{figure}
\epsscale{1.2}
\plotone{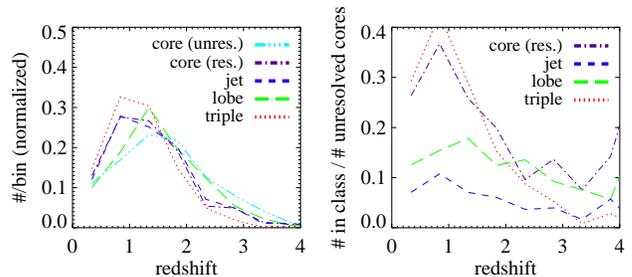}
\figcaption{\label{fig:redshifts}
Redshift distributions for the radio quasars according to morphology.  \emph{Left panel:} Sources in each class as a function of redshift.  \emph{Right panel:} Ratio of {\it resolved core}, {\it jet}, {\it lobe}, and {\it triple} sources to {\it unresolved core} sources as a function of redshift.}
\end{figure}

\subsection{Fraction of highly-reddened quasars as a function of radio morphology}
\label{subsec:color excess}

Previous observations of quasar colors suggest that radio quasars have redder SDSS colors than radio-quiet quasars \citep{i02}.  SDSS quasar targets are identified by selecting outliers from the stellar locus in SDSS color space \citet{richards02}.  Quasars with significant dust-reddening may lie on the stellar locus, and thus are not targeted by the color criterion.  However, {\it every} point source brighter than $i=19.1$ is targeted if it has a FIRST counterpart within 2\arcsec.  Thus, it has been suspected that the difference in colors between radio and non-radio quasars is merely a selection bias \citep{i02}.

With our morphology classifications in hand, we can now examine these color differences in more detail, and quantify the optical color biases in the SDSS quasar sample.  For example, systematic color differences among the radio morphology classes would indicate a physical cause, because all were targeted with the same algorithm.  However, if some radio quasar classes have the same color distribution as the radio-quiet sources, then it is unlikely that the colors of the radio-quiet quasars are strongly biased blue.  

\begin{figure}
\epsscale{1.15}
\plotone{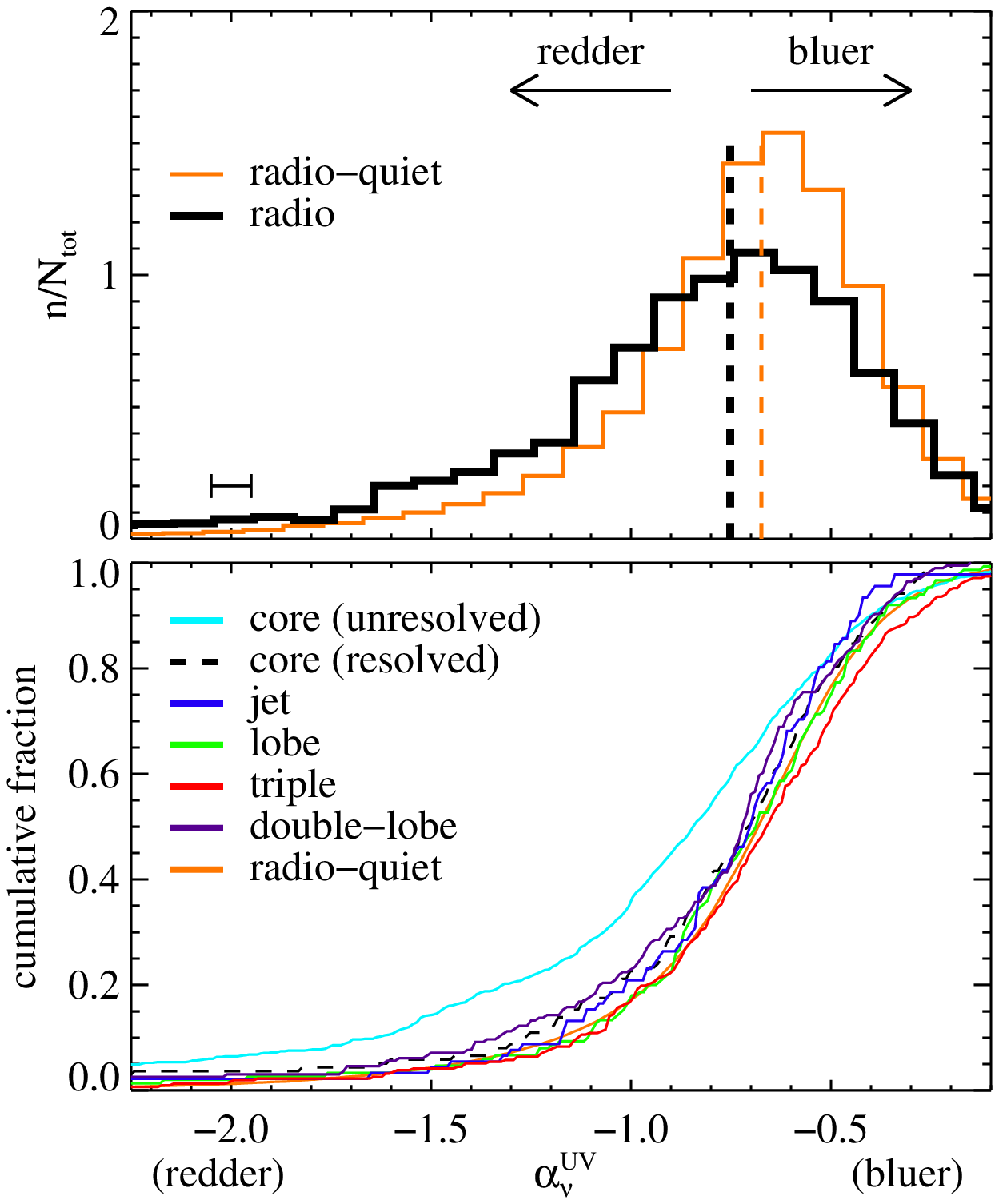}
\figcaption{\label{fig:alphahists}
Distributions of ultraviolet spectral index \alphaUV; more negative values indicate a redder continuum.  Top: Normalized histograms of \alphaUV\ for radio-quiet quasars (orange) and radio quasars (black) for sources with $i<19.1$.  Spectral index was calculated from $g-i$ color-excess (see text).  Dashed lines show the median values for each sample.  The error bar in the lower left shows the median scatter of \alphaUV\ about the $\delta (g-i)$ versus \alphaUV\ correlation.  Bottom: Cumulative distributions for individual radio morphology classes and the radio-quiet sample.}
\end{figure}

\begin{deluxetable}{cr}
\tablewidth{2in}
\tablecaption{Results of KS tests comparing \alphaUV\ distributions of radio quasars and radio-quiet quasars\label{table:ks}}
\tablehead{\colhead{Radio morphology} & \colhead{$P$-value}}
\startdata
unresolved core & 3.8e-35 \\
resolved core & 0.36 \\
jet & 0.52 \\
lobe & 0.54 \\
triple & 0.078 \\
double-lobe & 0.0051 \\
\enddata
\tablecomments{$P$-values indicate the probability that the radio quasar class has the same parent distribution as the radio-quiet class.}
\end{deluxetable}

We compare quasar ultraviolet spectral index ($\alpha_\nu$) distributions, as determined with the method of \citet{richards03}, who showed that the photometric spectral index of quasars correlates strongly with the photometric $g-i$ color.  They defined the $\Delta(g-i)$ ``color excess" as the difference between a quasar's $g-i$ and the median quasar $g-i$ at the same redshift.  The color excess correlates with ultraviolet spectral index, \alphaUV, as
\begin{equation}
\label{eq:alpha_nu}
\alpha^\mathrm{UV}_\nu = -0.6-2\times\Delta(g-i)
\end{equation}
(G. Richards, private communication); the correlation is not a function of redshift.  The median scatter in \alphaUV\ about the correlation is approximately 0.1.

The top panel in Figure~\ref{fig:alphahists} compares the \alphaUV\ distributions of radio-quiet quasars and radio quasars brighter than $i=19.1$, reproducing the \citet{i02} result with a much larger sample.  The long tails toward negative values indicate sources with extremely red spectra.  The radio quasars are shifted toward redder colors: the radio-quiet sample has a median value of $\alpha^\mathrm{UV}_\nu=-0.67$ and the radio sample has a median value of $\alpha^\mathrm{UV}_\nu=-0.75$.

The lower panel of Figure~\ref{fig:alphahists} shows cumulative distributions of \alphaUV\ for the individual radio morphology classes, including the double-lobed quasars.  For sources that have core radio emission, we apply a limit in radio flux density of 9.1~mJy \citep[14.0 mag, using the AB magnitude system of][]{okeGunn}; this is the limit above which the resolved/unresolved separation for the core sources is most effective (KI08).  The main result of this analysis is that unresolved core quasars have a significantly higher fraction of extremely red sources than all other quasar classes.  For example, unresolved cores make up 80\% of the radio sample with $\alpha^\mathrm{UV}_\nu<-1.5$, but only 45\% of the radio sample with $\alpha^\mathrm{UV}_\nu>-0.5$.

To quantify the statistical significance of differences in \alphaUV\ distributions for radio quasars and radio-quiet quasars, we employ the Kolmogorov-Smirnov (KS) test.  The $P$-value resulting from such a test represents the probability of obtaining the observed distributions by chance, if the two samples share the same parent distribution.  Table~\ref{table:ks} shows the results of KS tests comparing the radio-quiet quasar distribution to each radio class.  The KS tests quantify what is shown qualitatively in Figure~\ref{fig:alphahists}: the only quasar class with an extremely different ($P \ll 0.01$) color distribution from the radio-quiet quasars is the unresolved core class.  This difference must be a physical effect because unresolved core sources were targeted with the same algorithm as resolved cores, triples, lobes, and jets.  Because unresolved core quasars dominate the radio sample, a comparison of color distributions for the radio quasars and the radio-quiet quasars shows the difference reported by \citet{i02}.  Double-lobed quasars were targeted using the same criteria as radio-quiet quasars.  Because those two classes have not dissimilar observed color distributions, it is possible that they were drawn from similar parent color distributions.

The simplest interpretation of our results is: (1) unresolved core quasars are the only subsample of radio quasars that has an intrinsically redder color distribution than other quasars, and (2) optical colors of SDSS quasars are not strongly ($<0.1$~mag) biased blue.  We note that these results are specific to quasars in the redshift range $z\lesssim2.7$; quasars in Figure~\ref{fig:alphahists} are roughly limited to that range because of the redshift distribution of the S07 quasar catalog.  

It is not obvious why the unresolved cores have a much higher fraction of extremely red sources than the other samples.  Figure~\ref{fig:redshifts} suggests that this result is not a redshift effect, because lobe and jet sources have similar redshift distributions to the unresolved core sample, while triples and resolved cores have similar redshift distributions to each other.  As discussed in \citet{richards03}, dust-reddened spectra are not easily characterized by a single power law.  Our current analysis does not distinguish between spectra that are intrinsically red and those that are dust-reddened.  We do not expect that the highly reddened sources in the unresolved core sample are caused by dust reddening, because these are sources that should have the smallest viewing angles and thus the least obscuration by the AGN torus.

\section{COMPARISON BETWEEN TWO CORE-BOOSTING PARAMETERS, $R$ and \ri}
\label{sec:core-boosting}

Recent arguments suggest that \ri\ is a stronger statistical measure of core-boosting, and therefore of orientation, than $R$ (see discussion in \S\ref{sec:intro}).  $R$ normalizes the core boosting using the extended radio emission, which depends on age and environment in addition to orientation.  These factors should have less influence on \ri, which uses the core's optical luminosity to normalize the core boosting.  While a beamed synchrotron component (such as in blazars and BL Lac objects) would contaminate the \ri\ estimates in face-on sources, the presence of broad lines in the spectra (required for inclusion in the SDSS quasar sample) suggests that such a component is not seen in our sources.  In other words, it is unlikely that we have any directly face-on sources, as such orientations should result in an absence of any emission lines.

We now compare values of $R$ and \ri\ measured from our sample.  If the two parameters were both perfect measures of orientation, then they would correlate perfectly.  However, factors that affect $R$ and \ri\ separately, such as environment or age, can induce scatter in the correlation.

\begin{figure}
\epsscale{1.2}
\plotone{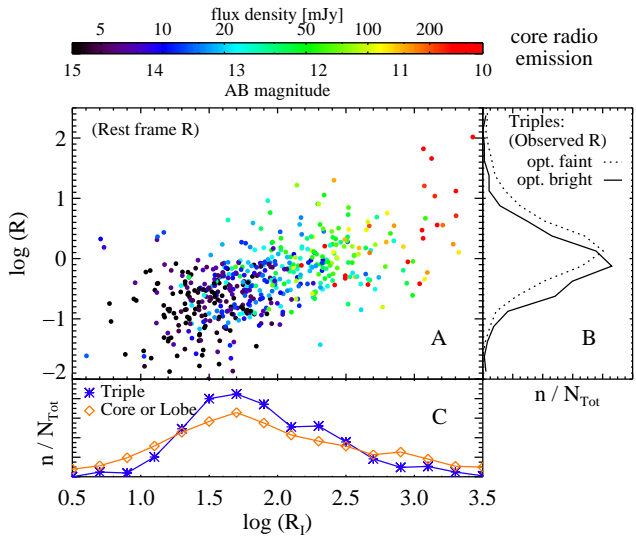}
\figcaption{\label{fig:rri_rcl_triple}
Correlation of $R$ and \ri\ for 519 {\it triple} quasars with well-determined core and lobe flux densities.  Panel A: Symbols are color-coded according to 20 cm core flux density.  A magnitude of 15 corresponds to 3.63 mJy; a magnitude of 10 corresponds to 363 mJy.  Panel~B: $R$ distributions of optically-bright and optically-faint triples, limited to sources with single-component FIRST detection of each lobe (see text).  Solid line shows the {\it observed-frame} $R$ distribution of points in panel~A; dotted line shows observed-frame $R$ for optically-faint triples.  Panel~C: \ri\ distribution of optically-bright triples (blue asterisks) and of the core and lobe sources (orange diamonds).}
\end{figure}

\subsection{Core-boosting parameters}

We define $R$ as the core-to-lobe flux density ratio at 20 cm, converting to the rest frame according to the K-correction formula
\begin{equation}
\label{eq:rcl}
\mathrm{R} = \mathrm{R_{[20\,cm]}} = \frac{S_\mathrm{core}}{S_\mathrm{lobe}}(1+z)^{(\alpha_\mathrm{core}-\alpha_\mathrm{lobe})},
\end{equation}
where $S_\mathrm{core}$ and $S_\mathrm{lobe}$ refer to observed 20~cm flux densities.  We assume a core spectral index of $\alpha_\mathrm{core}=0$ and a lobe spectral index of $\alpha_\mathrm{lobe}=-0.8$ (KI08).  

The radio-to-optical ratio is typically referred to in the literature as $R_\mathrm{V}$, the ratio of radio to visible light.  For our purposes, we use the ratio of core 20 cm luminosity to the luminosity in the $i$ band, which plays a large role in SDSS quasar selection.  Formally, the $i$-band, with an effective wavelength of 7481\AA, is slightly redder than what is typically considered the long-wavelength boundary of the optical regime at 7000\AA.  We refer to this ratio as \ri, defined as
\begin{equation}
\log(R_\mathrm{I}) = \log\left(\frac{L_\mathrm{radio}}{L_\mathrm{optical}}\right) = \frac{M_\mathrm{radio}-M_i}{-2.5},
\end{equation}
where $M_i$ is the Galactic reddening corrected \citep{sfd} and K-corrected $i$-band absolute magnitude from S07, who adopted a spectral index of $\alpha_\mathrm{opt}=-0.5$ to perform the K-correction.  $M_\mathrm{radio}$ is the analogous ``radio absolute magnitude" at 20~cm, obtained according to the formula
\begin{equation}
M_\mathrm{radio} = m_\mathrm{core} - k - D,
\end{equation}
where $D$ is the distance modulus and $k$ represents the K-correction \citep{kcorrect}:
\begin{equation}
k = -2.5 \times (1+\alpha_\mathrm{core}) \times \log(1+z).
\end{equation}
The ``radio apparent magnitude" $m_\mathrm{core}$ is the observed 20~cm flux density converted to apparent magnitude on the AB$_\nu$ system of \citet{okeGunn}, 
\begin{equation}
m_\mathrm{core} = -2.5\log\left(\frac{S_\mathrm{core}}{3631~\mathrm{Jy}}\right).
\end{equation}
As before, we adopt $\alpha_\mathrm{core}=0$.

\subsection{Evidence of a physical correlation between $R$~and~\ri}

Figure~\ref{fig:rri_rcl_triple} shows that $R$ and \ri\ are strongly correlated.  It also demonstrates that quasars with faint observed radio cores (indicated by symbol colors) have low core-boosting parameters, and vice-versa.  The relationship between core brightness and core-boosting parameter is most likely a selection effect.  To be classified as a triple, both of a quasar's lobes must be above the FIRST detection limit.  Therefore, triples with high $R$ must have cores that are much brighter than the FIRST limit.

We now investigate whether the $R$---\ri\ correlation is a selection effect or an intrinsic property of quasars.  For example, selection biases against high-\ri---low-$R$ sources (lower-right corner of panel~A) or against low-\ri---high-$R$ sources (upper-left corner of panel~A) would induce a spurious observed correlation.  The evidence presented in panels~B and~C, discussed below, suggests that the correlation is a true physical effect.

Panel~B indicates that the dearth of sources in the lower-right corner of panel~A is due to intrinsic quasar properties, for the following reasons.  At a given radio core flux density, sources with high \ri\ are more likely to fall below the faint limit for SDSS spectroscopic targeting; we refer to these sources as ``optically-faint triples", in contrast to ``optically-bright triples" in our main sample.  Optically-faint triples having lower $R$ values than optically-bright triples would be an indication that the empty lower-right corner in panel~A is due to a selection bias.  Panel~B compares the $R$ distributions of optically-bright and optically-faint triples.  The creation of the latter sample is described in detail in the Appendix.  For simplicity of sample selection, we require the optically-faint triples to have each of their two lobes detected as a single component in FIRST.  Therefore, in order to compare two equivalent data sets,  panel~B is limited to single-component lobes for the optically-bright triples as well.  For the optically-faint triples, we report $R$ values in the observed frame because we do not know their redshifts.  Therefore, panel~B also includes the observed-frame $R$ distribution for optically-bright triples.  The distributions indicate that optically-faint triples have $R$ values just as high (actually, even higher) than the optically-bright triples.  We conclude that optically-faint triples would not populate the lower-right corner of panel~A, and that the absence of such a population represents the true properties of quasars.

The absence of quasars in the upper-left corner of panel~A (with high $R$ and low \ri) is also due to intrinsic quasar properties, as demonstrated in panel~C.  Quasars can be classified as triples only if they have two lobes with $S_\mathrm{lobe}>1$mJy; otherwise, one or both lobes would be undetected.  Because of the optical and radio detection limits of our sample, quasars with high $R$ and low \ri, if they exist, are likely to have (an) undetected lobe(s), and therefore be classified as {\it core} or {\it lobe} rather than {\it triple}.  Panel~C compares the \ri\ distribution of the triples to that of the lobe plus core classes.  The latter do not show a lower range of \ri, which indicates that no such population of quasars with high $R$ and low \ri\ exists.

In summary, Figure~\ref{fig:rri_rcl_triple} shows a strong correlation between $R$ and \ri.  We demonstrated, using distributions in Panels~B and~C, that the observed correlation is not a selection effect, but is instead an intrinsic property: quasars with high $R$ tend to have high \ri, and vice-versa.  This observation supports the hypothesis that both parameters are measures of quasar orientation.  The scatter in the correlation supports the idea that other factors are also influencing these two measurements.

\section{THE DEPENDENCE OF OPTICAL SPECTRA ON RADIO SELECTION}
\label{sec:spectra}

The spectral composite is a valuable tool for studying properties of large samples of astronomical sources.  In recent years, the advent of large spectroscopic surveys has allowed for increasingly detailed studies of AGN spectral composites \citep{francis91,bakerHunstead95,zheng97,brotherton01,vandenberk01}.  Spectra of many objects can be combined to create composites with high signal-to-noise, allowing average properties to be compared across sample types.  Here we compare spectral properties of quasar radio morphology classes, and investigate whether a selection based on radio properties implies concomitant changes in optical spectra.

\begin{figure}
\epsscale{1.2}
\plotone{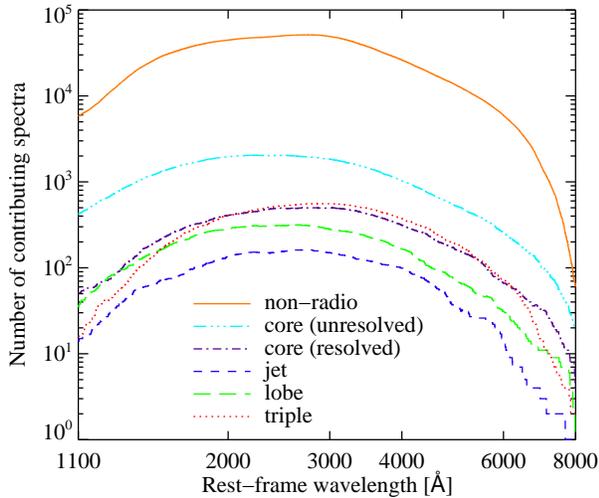}
\figcaption{\label{fig:numbers}
Number of spectra contributing to each spectral composite as a function of wavelength bin.  Colors correspond to the colors of spectral composites in Figure~\ref{fig:morph_radio}.}
\end{figure}

\subsection{Construction of quasar spectral composites}
\label{subsec:composites}

Creating a spectral composite involves several steps: shifting each input spectrum to its rest frame, rebinning the spectra to matching wavelength grids and resolution, normalizing their fluxes, and stacking into a final composite.  Variations of these steps can lead to significant differences in the resulting composite \citep{francis91}.

We redshift-corrected each spectrum using the S07 redshift (Section~\ref{subsec:sdss}).  Spectra were then re-binned, conserving flux, onto a common wavelength grid with bins $\sim69$~km~s$^{-1}$ wide, the same sampling as observed SDSS spectra.  

We used the \citet{vandenberk01} quasar spectral composite as a template for rescaling our input spectra to a common flux level, using the following method.  In the region of wavelength overlap, we found the flux ratio of the input spectrum to the Vanden Berk et al. composite at each wavelength; however we ignored the region blueward of the Ly$\alpha$ emission line ($\lambda<1250$\AA) so as to avoid uncertainties due to redshift-dependent absorption in the Ly$\alpha$ forest region.  We then determined the median flux ratio in the overlap region, and used this value to rescale the input spectrum.  Our resulting composites are in arbitrary flux density units (but on the same scale as the Vanden Berk et al. composite).

We stacked the rescaled spectra into two different final composites: a ``median" composite and a ``geometric mean" composite.  (For the latter, negative flux values were ignored.)  A median composite preserves the relative fluxes of emission features, while a geometric mean preserves the overall continuum shape for power-law spectra \citep{vandenberk01}.  We note that our composites do not include BAL quasars, as BALs can influence a spectrum's shape.

We compared the median composite for the 72,223 quasars (both radio and radio-quiet) without BALs to the median composite from Vanden Berk et al., which also excludes BAL quasars.  The two spectra agree very well and are nearly indistinguishable to the eye, except at the extreme wavelength values, where fewer spectra contribute and thus the composites are noisier.

The spectral composites for the entire sample and for individual morphological classes are discussed in the remainder of this section.  Table~\ref{table:composites_main} presents the composite for the radio-quiet quasars, for the full sample of radio quasars, and for the radio morphology classes.

\begin{deluxetable*}{r|rr|rr|rr|rr|rr|rr|rr}
\tablewidth{6in}
\tablecaption{\label{table:composites_main}Median Composite Quasar Spectra for Radio Morphology Classes}
\tablehead{\colhead{$\lambda$} & \multicolumn{2}{c}{Radio-quiet} & \multicolumn{2}{c}{All Radio} & \multicolumn{2}{c}{Res. Core} & \multicolumn{2}{c}{Unres. Core} & \multicolumn{2}{c}{Jet} & \multicolumn{2}{c}{Lobe} & \multicolumn{2}{c}{Triple} \\
\colhead{(\AA)} & \colhead{$f_\lambda$} & \colhead{$\sigma_f$} &
\colhead{$f_\lambda$} & \colhead{$\sigma_f$} & \colhead{$f_\lambda$} & \colhead{$\sigma_f$} & \colhead{$f_\lambda$} & \colhead{$\sigma_f$} & \colhead{$f_\lambda$} & \colhead{$\sigma_f$} & \colhead{$f_\lambda$} & \colhead{$\sigma_f$} & \colhead{$f_\lambda$} & \colhead{$\sigma_f$}}
\startdata
900.00 & 1.80 & 0.148 & 1.90 & 0.681 & 0.664 & 1.65 & 2.18 & 0.822 & 1.39 & 3.78 & 2.66 & 1.34 & 4.40 & -1.00 \\
900.21 & 1.86 & 0.149 & 2.19 & 0.693 & 0.980 & 1.58 & 2.47 & 0.844 & 1.83 & 1.80 & 2.22 & 1.78 & 5.37 & -1.00 \\
900.41 & 1.77 & 0.144 & 1.80 & 0.713 & 0.808 & 1.38 & 1.81 & 0.847 & 0.488 & 4.16 & 3.57 & 2.68 & 4.96 & -1.00 \\
900.62 & 1.90 & 0.143 & 1.76 & 0.740 & 0.880 & 2.40 & 2.09 & 0.847 & 1.13 & 3.01 & 1.19 & 2.83 & 2.28 & -1.00 \\
900.83 & 1.88 & 0.140 & 1.95 & 0.798 & 1.28 & 1.75 & 1.78 & 0.921 & 2.83 & 2.76 & 2.06 & 4.22 & 3.79 & -1.00 \\
\enddata
\tablecomments{The full table is available in the electronic version of the paper.  Values of $f_\lambda$ (flux) are in arbitrary units, such that they are on the same flux scale as the \citet{vandenberk01} spectrum (see \S\ref{subsec:composites} for details).  The uncertainty in the median is given by $\sigma_f$;   a value of $-1$ indicates that only one input spectrum contributes to that wavelength bin, while 0 indicates that no input spectra contribute at that wavelength.}
\end{deluxetable*}

Individual quasar spectra have different rest-frame wavelength coverage owing to their wide redshift range ($0.78<z\lesssim5.4$).  For each spectral composite, different objects contribute to the long-wavelength and short-wavelength regions.  Figure~\ref{fig:numbers} shows the number of input spectra that make up each section of a spectral composite as a function of wavelength.

\subsection{Spectral composites of individual radio morphology classes}

\begin{figure}
\epsscale{1.2}
\plotone{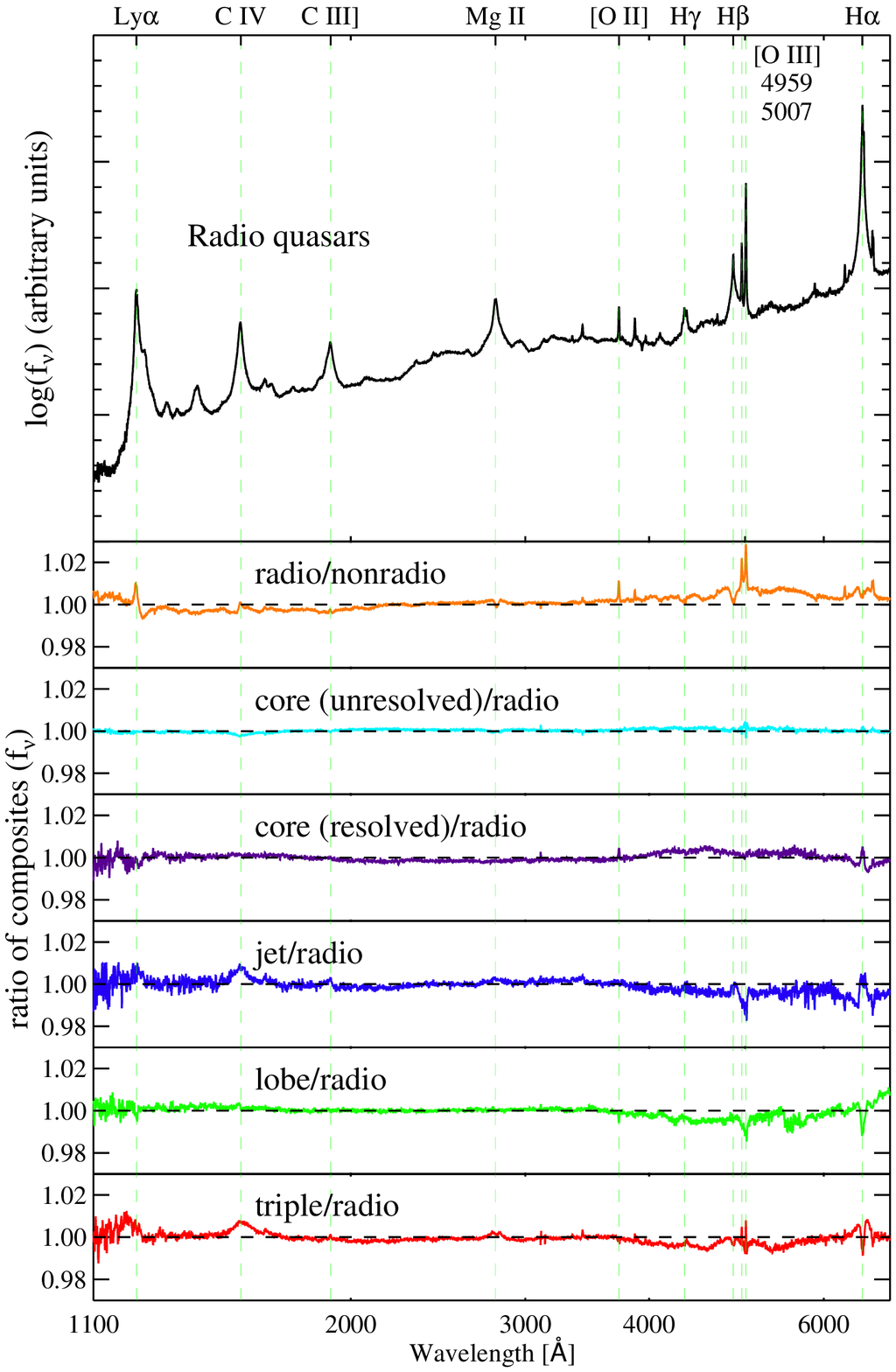}
\figcaption{\label{fig:morph_radio}
Comparison of geometric mean spectral composites of radio quasars according to morphology.  Top panel: composite spectrum for radio quasars.  Second panel: ratio of radio quasar composite to composite for radio-quiet quasars.  Third panel: ratio of composite for unresolved core sources to all radio sources.  The remaining four panels show analogous ratios for resolved cores, jets, lobes, and triples compared to the full radio sample.}
\end{figure}

\begin{figure}
\epsscale{1.2}
\plotone{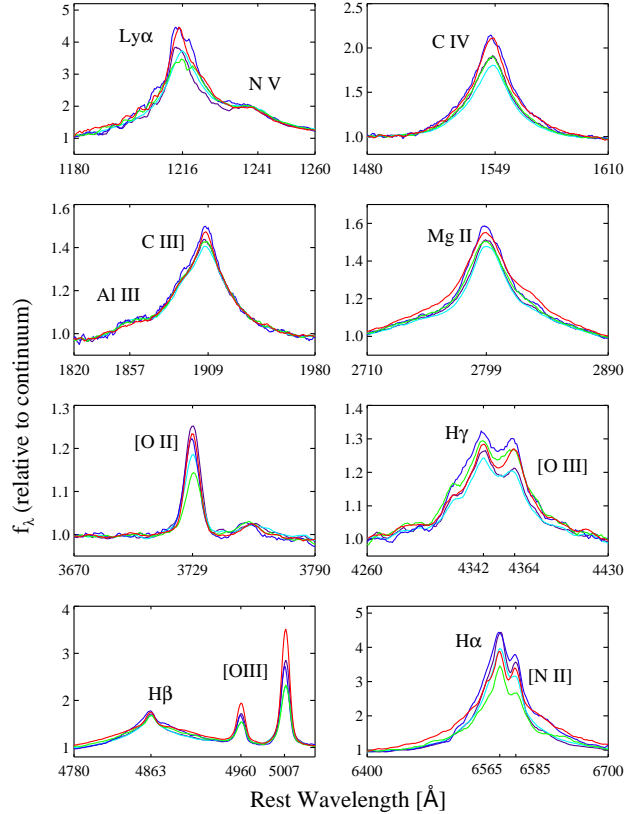}
\figcaption{\label{fig:speclines_radio}
Spectral line profiles from the median composite spectra of each radio morphology class.  Colors correspond to the colors in Figure~\ref{fig:numbers}: cyan---unresolved core; purple---resolved core; blue, jet; green, lobe; red, triple.  The spectral line profiles were fit to a simple Gaussian + line continuum model, and have been rescaled such that the continuum at the line center is equal to 1.}
\end{figure}

\begin{deluxetable*}{r|cc|cc|cc|cc}
\tablewidth{5in}
\tablecaption{\label{table:composites_triple}Median Composite Quasar Spectra for Triple Quasar Log($R$) Subclasses}
\tablehead{\colhead{$\lambda$} 
& \multicolumn{2}{c}{log($R$) $<-1.5$}
& \multicolumn{2}{c}{$-1.5<$ log($R$) $<-0.5$} 
& \multicolumn{2}{c}{$-0.5<$ log($R$) $<0.5$} 
& \multicolumn{2}{c}{$0.5<$ log($R$)} \\
\colhead{(\AA)} & \colhead{$f_\lambda$} & \colhead{$\sigma_f$} & \colhead{$f_\lambda$} & \colhead{$\sigma_f$} & \colhead{$f_\lambda$} & \colhead{$\sigma_f$} & \colhead{$f_\lambda$} & \colhead{$\sigma_f$}}
\startdata
920.14 & 24.3 & -1.00 & 0.00 & 0.00 & 0.00 & 0.00 & 0.00 & 0.00 \\
920.35 & 19.3 & -1.00 & 0.00 & 0.00 & 0.00 & 0.00 & 0.00 & 0.00 \\
920.56 & 6.94 & -1.00 & 0.00 & 0.00 & 0.00 & 0.00 & 0.00 & 0.00 \\
920.78 & 8.78 & -1.00 & 0.00 & 0.00 & 0.00 & 0.00 & 0.00 & 0.00 \\
920.99 & 6.27 & -1.00 & 0.00 & 0.00 & 0.00 & 0.00 & 0.00 & 0.00 \\
\enddata
\tablecomments{The full table is available in the electronic version of this paper.  Values of $f_\lambda$ (flux) are in arbitrary units, such that they are on the same flux scale as the \citet{vandenberk01} spectrum (see \S\ref{subsec:composites} for details).  The uncertainty in the median is given by $\sigma_f$;   a value of $-1$ indicates that only one input spectrum contributes to that wavelength bin, while 0 indicates that no input spectra contribute at that wavelength.}
\end{deluxetable*}

Geometric mean composites for distinct radio morphology classes are compared by ratio in Figure~\ref{fig:morph_radio}.  Differences in individual composites are difficult to distinguish by eye, but are emphasized in {\it ratios} of the composites.  We have separated the core sources into resolved and unresolved (\S\ref{subsubsec:redshift effects}).  Because those two groups dominate the radio quasars, the core/radio ratios are nearly equal to 1 at all wavelengths.  Composites for different morphology classes show similar spectral shapes.  The second panel suggests that the average optical spectral index  is redder for radio quasars than for radio-quiet quasars (see \S\ref{subsec:color excess} for further discussion).  However, spectral indices measured from composites may not be accurate because objects at different redshifts contribute at different wavelengths owing to varied rest-frame wavelength coverage.  Thus the spectral index of a composite depends on quasar evolution.

The ratios of spectral composites show that the strength of spectral lines vary with radio power and morphology.  For example, \civline\ is strongest in jet and triple sources.  Figure~\ref{fig:speclines_radio} directly compares the line profiles of the median spectral composites.  Spectra in this figure have been scaled such that the continuum value at the line center is equal to~1.

In Figure~\ref{fig:morph_radio}, the [O\,III] doublet and [O\,II] are clearly stronger in radio quasars than radio-quiet quasars, an effect that is expected from previous observations of these lines.  \citet{rawlingsSaunders91} observed that the total narrow line luminosity in quasars (measured from [O\,II] and [O\,III]) positively correlates with jet power, as measured from the lobe emission and spectral ages.  In our sample, these lines are strongest in the triples and weakest in the lobes and jets.  These results agree with Rawlings \& Saunders, given that triples have greater extended emission than the lobe and jet sources (which suggests that triples have the most powerful radio jets).  The resolved cores are weaker in [O\,III] than the triples, but stronger than the lobes and jets, and are even stronger in [O\,II] than the triples.  This evidence suggests that the resolved cores contain many powerful sources.  As surmised in \S\ref{subsec:morphology}, these sources are probably still young, and therefore small in physical extent, which explains why their lobes are not resolved into separate components in FIRST.

Among the broad lines, C\,III] and Mg\,II show no significant change among the radio classes.  However, Ly$\alpha$ and C\,IV show strong variation, and are clearly the strongest in the triple and jet samples.  

\subsection{The effects of orientation on composite spectra}
\label{subsec:orientation_composites}

By dividing the sample of triples according to $R$, we can create spectral composites for quasars with different ranges of assumed orientation angle to the radio jet axis.  \citet{bakerHunstead95} performed a similar analysis of 47 radio quasars.  Their results suggest that as the viewing angle to the jet axis increases, the optical continuum steepens and line widths increase.  We now revisit these findings using an order-of-magnitude larger sample.

We divided the triples into the following four groups based on the rest-frame core-to-lobe ratio $R$ at 20~cm (Eq.~\ref{eq:rcl}): (1) $0.5<$ log($R$), (2) $-0.5<$ log($R$) $<0.5$, (3) $-1.5<$ log($R$) $<-0.5$, (4) log($R$) $<-1.5$.  These groups contain 30, 277, 190, and 20 triples, respectively.  Groups 2, 3, and 4 roughly correspond to the three groups defined by \citet{bakerHunstead95}, who used a measure of $R$ at 3~cm in the rest frame.  The conversion between the 3~cm and 20~cm values is
\begin{equation}
\mathrm{R_{[20\,cm]}} = (20/3)^{-(4/5)}\times\mathrm{R_{[3\,cm]}}.
\end{equation}
Based on the quasar core-boosting models, group~1 should correspond to quasars with lines-of-sight close to the radio-jet axis, while group~4 should contain quasars viewed closest to the plane of the accretion disk.  Group~4 contains only 20 sources from the main radio quasar sample.  We add to this category by including double-lobed quasars, which are clearly lobe-dominated by virtue of having no radio-detected core.  We calculated an upper limit on $R$ for the double-lobed quasars, assuming a core flux density less than the FIRST survey limit of 1~mJy.  Ninety-seven double-lobed quasars have an upper limit of log($R$) $<-1.5$, and we included their spectra when making the composite spectrum for group~4.  The median spectral composites for the four log($R$) groups are presented in Table~\ref{table:composites_triple}.

\begin{figure}
\epsscale{1.2}
\plotone{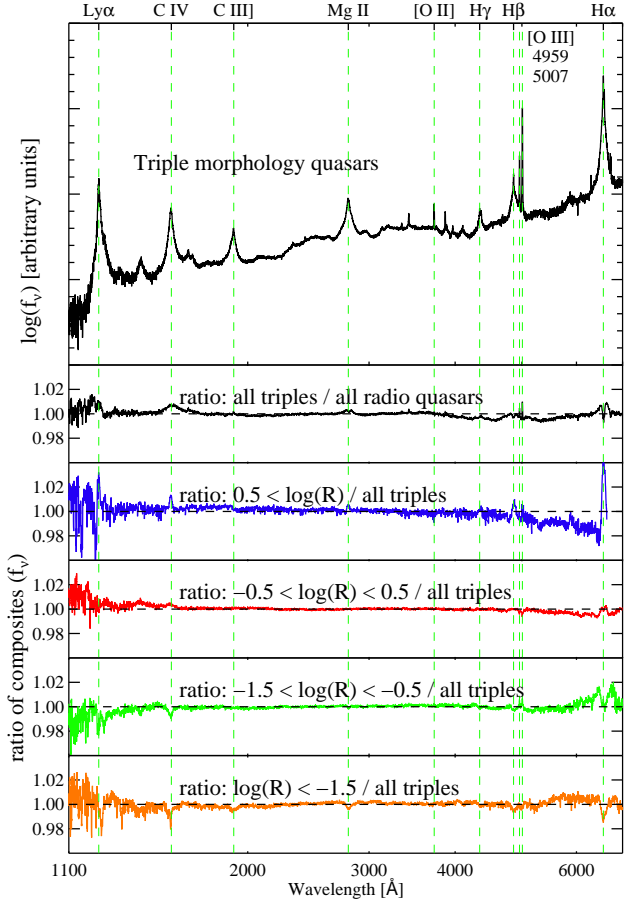}
\figcaption{\label{fig:morph_triples}
Comparison of geometric mean spectral composites for quasars categorized by core-to-lobe ratio at 20 cm in the rest frame.  Top panel: composite spectrum for all {\it triples}.  Second panel from top: ratio of {\it triple} composite to radio quasar composite.  Remaining panels: ratios of other spectral composites compared to the spectrum shown in the top panel.  The third, fourth, and fifth panels (blue, red, and green) correspond to subsets of the {\it triples} sample.  The bottom panel includes double-lobed quasars without a radio core.}
\end{figure}

The geometric mean composites are shown in Figure~\ref{fig:morph_triples}.  We compare them by examining the ratio of each with the composite spectrum for all triples.  Contrary to the results of \citet{bakerHunstead95}, we see no change in spectral shape with $R$.  However, there is a clear change in the \civline\ line profile: \ewciv\ appears to increase monotonically toward sources with the highest $R$.

\begin{figure}
\epsscale{1.23}
\plotone{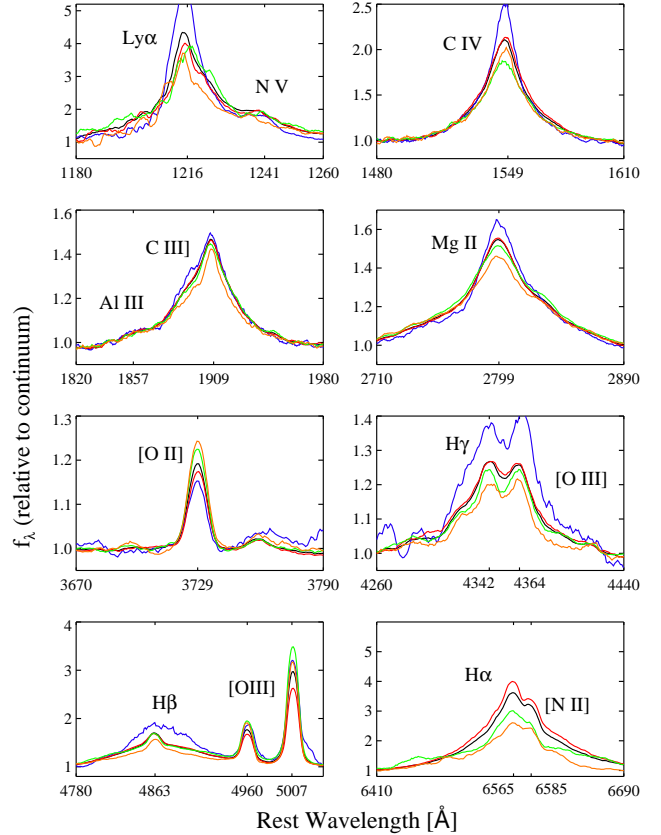}
\figcaption{\label{fig:speclines_triple}
Spectral line profiles from the median composite spectra of quasars categorized by core-to-lobe ratio at 20 cm in the rest frame.  The samples and colors in this figure correspond to those in Figure~\ref{fig:morph_triples}: black, all {\it triples}; blue, triples with log($R$) $\geq0.5$; red, triples with $-0.5\leq$ log($R$) $<0.5$; green, triples with $-1.5<$ log($R$) $<-0.5$; orange, triples with log($R$) $<-1.5$.  The spectral line profiles were fit to a simple Gaussian + line continuum model, and have been rescaled such that the continuum at the line center is equal to~1.}
\end{figure}

Individual line profiles are shown in Figure~\ref{fig:speclines_triple} using the median composites.  It appears that the EW of the broad lines (C\,IV, C\,III], Mg\,II and the Balmer lines) increase with $R$.  In contrast, the [O\,II] and [O\,III] lines show roughly the opposite trend: their EW appears to be anti-correlated with $R$.  Previous observations have suggested that both narrow and broad line EW are anti-correlated with $R$, as claimed for Mg\,II by \citet{browneMurphy87}, \citet{bakerHunstead95}, and \citet{baker97}; for the Balmer lines by \citet{bakerHunstead95}; and for [O\,III] by \citet{willsBrowne86}, \citet{jacksonBrowne91}, and \citet{baker97}.  These results from our composite spectra are consistent with previous claims for the narrow lines, but are inconsistent for the broad lines.  However, we note that the scatter of EW in individual spectra contributing to each composite is large, such that the uncertainty in the median EW for each line tends to be much greater than the change in EW seen from one composite spectrum to the next.  We will investigate these trends quantitatively in \S\ref{sec:line_profiles}; the results of correlation analyses presented there suggest that emission line EW is not significantly correlated with $R$. 

\begin{deluxetable*}{r|cc|cc|cc|cc}
\tablewidth{5in}
\tablecaption{\label{table:composites_logrri}Median Composite Quasar Spectra for Quasar Log(\ri) Subclasses}
\tablehead{\colhead{$\lambda$} 
& \multicolumn{2}{c}{log(\ri) $<1.5$}
& \multicolumn{2}{c}{$1.5<$ log(\ri) $<2.0$} 
& \multicolumn{2}{c}{$2.0<$ log(\ri) $<2.5$} 
& \multicolumn{2}{c}{$2.5<$ log(\ri)} \\
\colhead{(\AA)} & \colhead{$f_\lambda$} & \colhead{$\sigma_f$} & \colhead{$f_\lambda$} & \colhead{$\sigma_f$} & \colhead{$f_\lambda$} & \colhead{$\sigma_f$} & \colhead{$f_\lambda$} & \colhead{$\sigma_f$}}
\startdata
900.00 & 3.18 & 0.814 & 1.79 & 1.36 & 1.48 & 1.27 & 1.90 & 1.76 \\
900.21 & 1.88 & 0.938 & 2.19 & 1.12 & 1.83 & 1.48 & 2.90 & 1.93 \\
900.41 & 0.956 & 1.17 & 1.67 & 1.36 & 1.63 & 1.05 & 2.07 & 2.05 \\
900.62 & 1.25 & 1.11 & 2.83 & 1.30 & 2.02 & 1.50 & 1.28 & 1.96 \\
900.83 & 1.29 & 1.24 & 2.79 & 1.42 & 2.65 & 1.18 & 1.25 & 2.55 \\
\enddata
\tablecomments{The full table is available in the electronic version of this paper.  Values of $f_\lambda$ (flux) are in arbitrary units, such that they are on the same flux scale as the \citet{vandenberk01} spectrum (see \S\ref{subsec:composites} for details).  The uncertainty in the median is given by $\sigma_f$;   a value of $-1$ indicates that only one input spectrum contributes to that wavelength bin, while 0 indicates that no input spectra contribute at that wavelength.}
\end{deluxetable*}

\begin{figure}
\epsscale{1.2}
\plotone{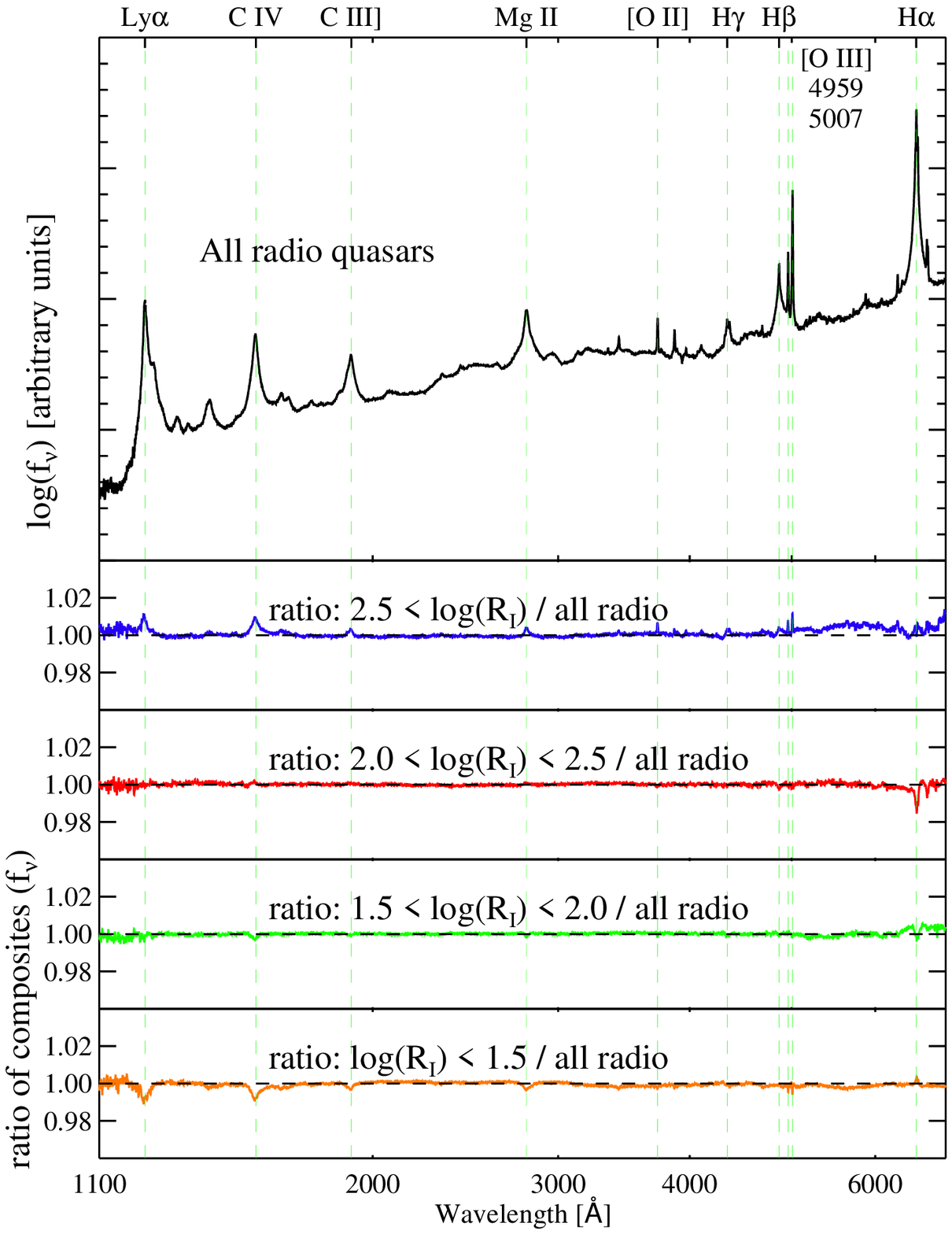}
\figcaption{\label{fig:morph_logrri}Comparison of geometric mean spectral composites for quasars categorized by log(\ri).  The top panel shows the composite spectrum for all radio quasars.  The remaining panels each show the ratio of a composite for a subset of quasars (as labeled) to the composite in the top panel.}
\end{figure}

\begin{figure}
\epsscale{1.23}
\plotone{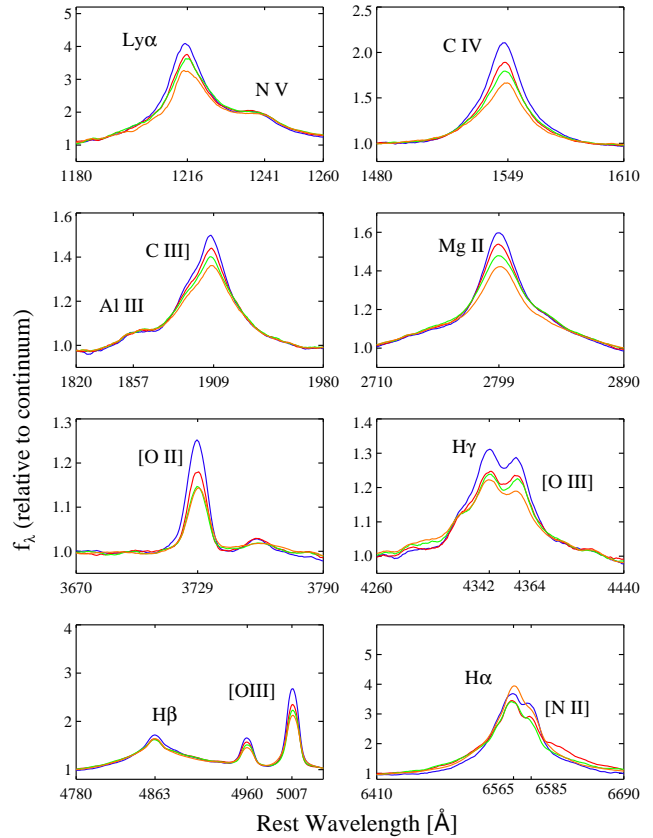}
\figcaption{\label{fig:speclines_logrri}Spectral line profiles from the median composite spectra of radio quasars categorized by core radio-to-optical flux density ratio in the rest frame.  The samples and colors correspond to those in Figure~\ref{fig:morph_logrri}: blue, log(\ri)$>2.5$; red, $2<\log($\ri$)<2.5$; green, $1.5<\log($\ri$)<2$; orange, $\log($\ri$)<1.5$. The spectral line profiles were fit to a simple Gaussian + line continuum model, and have been rescaled such that the continuum at the line center is equal to~1.}
\end{figure}

Motivated by our conclusion in Section~\ref{sec:core-boosting} that $R$ and \ri\ are physically correlated, and by evidence that the latter is the stronger statistical measure of orientation \citep{willsBrotherton95, kharbEtal10}, we now investigate composite spectra grouped by \ri.  We created the following four groups: (1) $2.5<$ log($R$), (2) $2.0<$ log($R$) $<2.5$, (3) $1.5<$ log($R$) $<2.0$, (4) log($R$) $<1.5$.  These groups contain 1070, 960, 1412, and 1272 quasars, respectively.  These groups are much larger than the log($R$) groups because log(\ri) can be determined for every quasar, not just for triples.

The geometric mean composites for the four \ri\ quasar groups are shown in Figure~\ref{fig:morph_logrri}.  Strong differences in Ly$\alpha$ and C\,IV are apparent in the ratios of composites, as well as subtle differences in C\,III], Mg\,II, and H$\beta$.  Each of these lines appears to be the strongest in group (1), with decreasing strength toward group (4).  These differences are even more clear in Figure~\ref{fig:speclines_logrri}, which shows the emission lines in the median spectral composites.  
This figure indicates that the EW of quasar emission lines increases as \ri\ increases.  In the following section, we investigate these intriguing trends more quantitatively by examining correlations among individual spectra.1

\section{CORRELATIONS OF SPECTRAL LINE PARAMETERS WITH $R$ AND \ri}
\label{sec:line_profiles}

Three commonly-accepted theories that relate quasar line behavior to viewing angle are: orientation-dependent obscuration by a clumpy dust torus \citep[e.g.,][]{nenkovaEtal08}, Doppler-boosting of the continuum \citep[e.g.,][]{browneMurphy87}, and inclination angle of the accretion disk \citet[e.g.,][]{netzer87}.  Each model predicts that the EW of quasar emission lines increases as the viewing angle to the radio-jet axis increases.  These theories share the assumptions that the line emission is isotropic while the continuum emission depends on the orientation to the accretion disk.

In this section, we examine the dependence of quasar emission lines on the core-boosting parameters $R$ and \ri.  We find no significant correlations with $R$, but find several significant correlations with \ri\ in the {\it opposite} direction from what is predicted by the above models.

\subsection{Measuring correlations in the data}

To evaluate correlations between line parameter and $R$ or \ri, we ran the linear regression code of \citet{kelly07}, which uses Bayesian methods to incorporate uncertainties in the data, as well as upper-limits.  The parameterization we used is 
\begin{equation}
\label{eq:fits}
\log(Y) = A + B \times\log(X),
\end{equation}
where $Y$ is the dependent variable (e.g., EW), $X$ is the independent variable (e.g., $R$ or \ri), and $A$ and $B$ are intercept and slope respectively.  All line parameters $X$ have 1-$\sigma$ error bars $\sigma_X$ determined from \citet{shenEtal10}.  For sources where $X-\sigma_X<0$, we treated the source as a non-detection with upper-limit of $X+3\sigma_X$.  This modification was necessary in approximately 1.5\% of sources. 

To investigate correlations with $R$, we used the sample of 519 triple quasars and 383 lobe quasars with known core and lobe flux densities (\S\ref{subsubsec:triples}).\footnote{The remaining 100 triple quasars and 4 lobe quasars were excluded due to crowded FIRST images.  The identities of their lobe components, and therefore their $R$ values, are ambiguous (\S\ref{subsubsec:triples}).}  To investigate correlations with \ri, we used the full radio quasar sample.  We limited the analysis to spectra with median signal-to-noise per pixel (SDSS pixels are 69 km s$^{-1}$ wide) of at least 5 in the relevant line region \citep{shenEtal10}.  We excluded sources with a broad absorption line feature near the line, as identified by \citet{gibson09_bal}; such features can affect the measurement of line EW.  We also excluded sources with the line in absorption.  Such exclusions do not hinder the Bayesian linear regression analysis, as discussed in \citet{kelly07}.

In addition to estimating the intercept $A$ and slope $B$, the code estimates the Pearson correlation coefficient.  The latter ranges from $-1$ to 1 where 0 indicates no correlation, 1 indicates a perfect correlation, and $-1$ indicates a perfect anti-correlation.  We consider a correlation to be significant if both the slope $B$ and the correlation coefficient are non-zero with certainty $\geq3\sigma$.  Results of each test are listed in Table~\ref{table:fits}, and are discussed qualitatively in the remainder of this section.

The scatter in EW measurements is much larger than errors in individual data points.  A number of intrinsic processes may contribute to the scatter.  Quasars are known to vary in the optical continuum by about 20\% over multi-year timescales \citep[e.g.,][]{hookEtal94}, which may be due to changes in the accretion disk or jet variations \citep{marscher}.  As demonstrated by, e.g., \citet{vandenberk04}, variability amplitude in the optical depends on wavelength (quasars are more variable at shorter optical wavelengths), luminosity (more luminous quasars vary more strongly), and redshift (higher redshift quasars show stronger variability).  Radio variability of the core can occur at the 20\% level \citep{barvainis} although this is typically observed in only 1-2\% of FIRST sources \citep{rys,devries04}.  Differences in local environment may also cause variations in emission line luminosities \citep[e.g.,][]{mccarthy93}.

\begin{figure*}
\epsscale{1.15}
\plotone{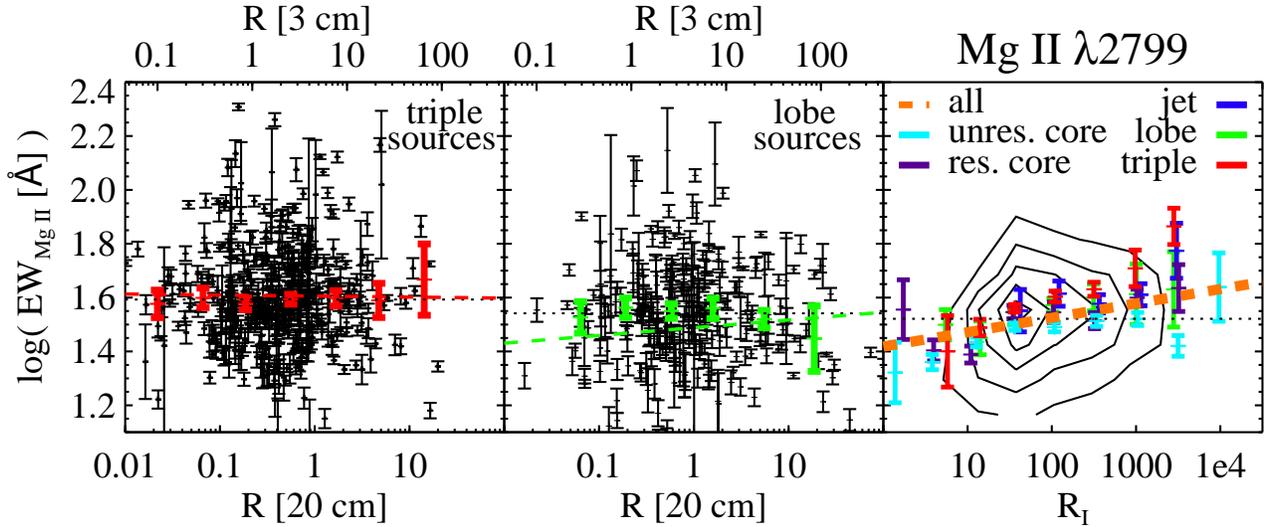}
\figcaption{\label{fig:linecorr MgII}
EW of Mg\,II $\lambda2799$ as a function of core-boosting parameters $R$ and \ri.  In the left and middle panels, the full distributions are shown as black points.  In the right panel, the full distribution is shown as linearly-spaced contours.  The number of sources in each panel and each morphology class is listed in Table~\ref{table:fits}.  To guide the eye, colored error bars show the median (and error in the median) EW in evenly-spaced log($R$) or log(\ri) bins.  Horizontal dotted lines show the median value for each panel.  Dashed lines show the best-fit line for the full set of sources in each panel.  Quantitative results of the linear regression for all morphology classes are listed in Table~\ref{table:fits}.}
\end{figure*}

\begin{figure*}
\epsscale{1.15}
\plotone{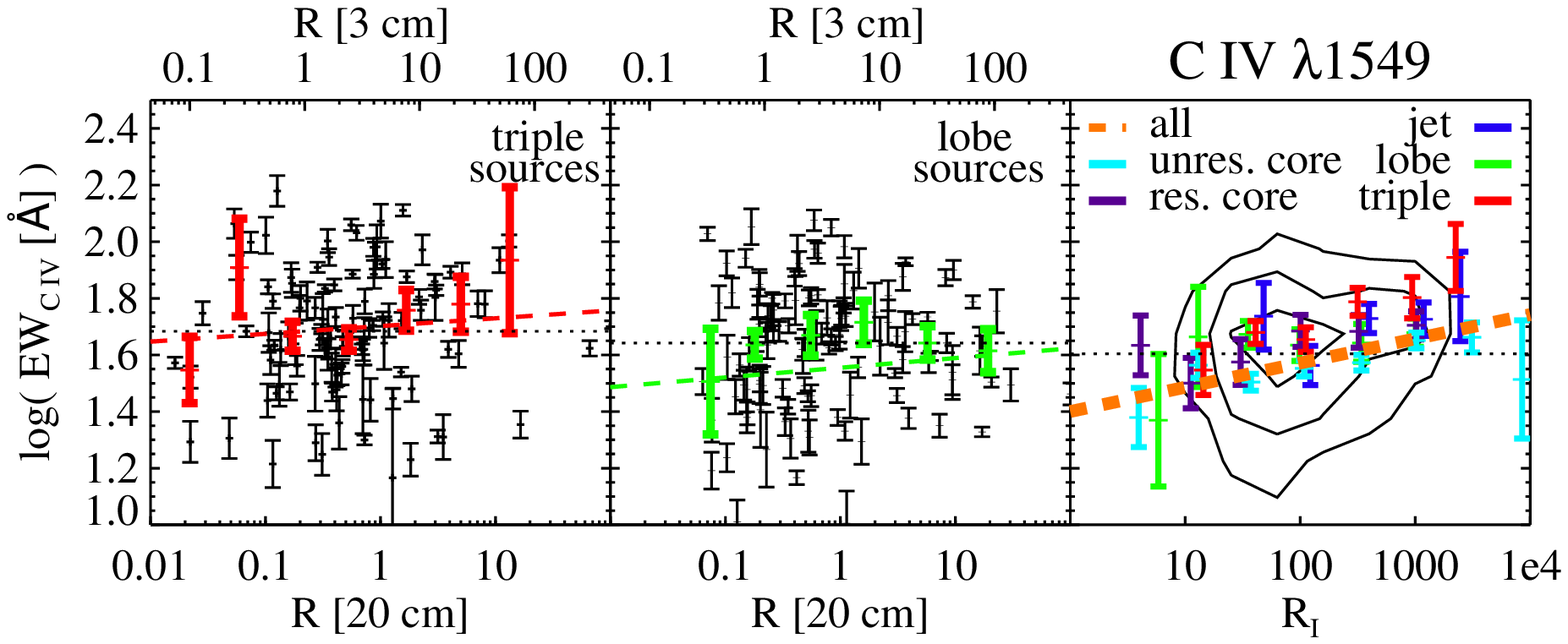}
\figcaption{\label{fig:linecorr CIV}
EW of C\,IV $\lambda1549$ as a function of core-boosting parameters $R$ and \ri.  Colors and symbols are as in Figure~\ref{fig:linecorr MgII}.  Number of sources of each class and results of the linear regression analysis are listed in Table~\ref{table:fits}.}
\end{figure*}

\subsection{Equivalent width of Mg\,II and C\,IV broad emission lines}
\label{subsec:EWline}

We begin by investigating the EW of Mg\,II $\lambda2799$ (\ewmgii) as a function of $R$ and \ri\ (Fig.~\ref{fig:linecorr MgII}).  The left panel of Figure~\ref{fig:linecorr MgII} shows \ewmgii\ vs $R$ for {\it triple}-morphology quasars, using the average lobe flux density to calculate $R$.  The middle panel shows \ewmgii\ vs $R$ for {\it lobe}-morphology quasars, using the single detected lobe to calculate $R$.  For plotting purposes only, we show median values of \ewmgii\ in $\Delta\log(R)=0.5$ bins to help guide the eye.  To perform the linear regression analysis, we used the true data points, {\it not} the median values.

There is no significant correlation of \ewmgii\ with $R$; this result is reflected quantitatively via the results of the linear regression (Table~\ref{table:fits}).  However, we do see a strong correlation with \ri\ in the rightmost panel.  In the full sample of radio quasars, the correlation is significant at approximately the 10$\sigma$ level.  When dividing the radio quasars sample into individual morphology classes, we continue to see a significant correlation, but with reduced strength ($\sim$4--7$\sigma$) in the core, triple, and jet classes.  The results of the linear regression vary across the different morphology classes; the triple and jet classes have the steepest slopes while the core-morphology sources show weaker, but still significant, correlations.  The lobe sources show the same trend, but not at a significant level (only 1.6$\sigma$).  The greatest variation is seen between the unresolved core class and the triple class, which differ in slope by $\sim3\sigma$.  We conclude that there is certainly a positive trend of \ewmgii\ with \ri\ in radio quasars, and that the trend may vary weakly with morphology class.

Figure~\ref{fig:linecorr CIV} shows the equivalent analysis for \ewciv.  Similar to the results for Mg\,II, we see no correlation of EW with $R$, but we do see a significant positive correlation ($\sim8\sigma$) with \ri.  The same trend with \ri\ is observed in the individual morphology classes, but at a weaker level.  We observe significant correlations for the core and triple classes, but not for the lobe or jet classes.  In this case, the overall variation among the different classes is only $\sim1.2\sigma$.

Our observations conflict with the results of \citet{baker97}.  Baker observed a significant anti-correlation of \ewmgii\ with $R$ in a sample of 67 quasars with $z<2.2$; a similar qualitative, but not significant, trend was observed for \ewciv.  Baker reported a stronger trend for \ewmgii\ in quasars with $z>1$ than in those with $z<1$.  We see no difference in correlations with $R$ when testing samples in these two redshift ranges.  Baker proposed orientation-dependent obscuration by a dusty torus as an explanation for their observed trend, suggesting that the torus increasingly obscures the accretion disk at high inclination angles.  We note that, while our quasars match the Baker sources in redshift (e.g., $z\lesssim2.2$ for Mg\,II emitters), our samples differ in radio brightness: the Baker sources had radio flux densities $S>0.95$ Jy at 408 MHz; the brightest Mg\,II triple in our sample has a radio core flux density of 0.71~Jy.  If orientation-dependent obscuration is responsible for the trend observed by Baker, our results may show that dust obscuration does not depend on orientation in these lower-luminosity quasars.  Alternatively, our results could indicate that the continuum emission and the broad line region (BLR) intrinsically have the same dependence on orientation such that EW does not change with orientation.  This situation could occur if both the broad line emission and the continuum emission originate in the accretion disk \citep[e.g.,][]{collin-souffrin87}.

Why is there a significant correlation of EW with \ri\ but not with $R$?  The \ri\ distribution may have a higher signal-to-noise ratio because it includes a larger number of data points: $R$ can only be determined for quasars with at least one observed radio lobe, while \ri\ can be determined for all radio quasars.  However, we observe a stronger \ri\ dependence even among the triple sources alone.  One possible explanation is that the EW--\ri\ correlation is based on some intrinsic physical cause, rather than on orientation.  However, if both parameters are statistical indicators of orientation, then our data suggest that broad line EW does depend on orientation, and that \ri\ is the stronger statistical orientation parameter.  This conclusion is consistent with the proposal of \citet{willsBrotherton95} and \citet{kharbEtal10} that $R$ additionally depends on local environment, while \ri\ does not.  One possible explanation for these \ri--EW correlations is an anisotropically-emitting BLR, composed of spherical clouds that emit most intensely from the inner side of each cloud, illuminated by the ionizing disk emission \citep[e.g.,][]{ferlandEtal92,koristaEtal97}.  An edge-on observer would see ionized emission only from clouds on the far side of the BLR for, e.g., C\,IV \citep[see Fig.~4 of][]{horneEtal03}.  In contrast, a face-on observer would see half the ionized face of every BLR cloud, resulting in more BLR emission relative to the continuum and therefore a larger EW.

\begin{figure*}
\epsscale{1.15}
\plotone{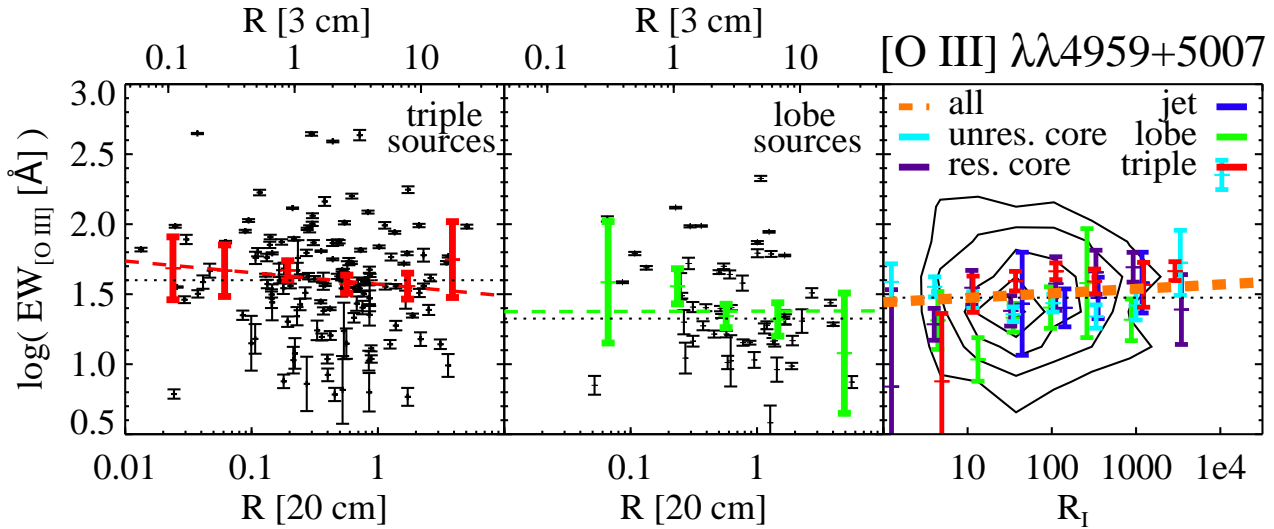}
\figcaption{\label{fig:linecorr OIII}
EW of [O\,III] $\lambda\lambda 4959/5007$ as a function of core-boosting parameters $R$ and \ri.  Colors and symbols are as in Figure~\ref{fig:linecorr MgII}.  Number of sources of each class and results of the linear regression analysis are listed in Table~\ref{table:fits}.}
\end{figure*}

\subsection{Equivalent width of the [O\,III] narrow line}

\begin{figure}
\epsscale{1.2}
\plotone{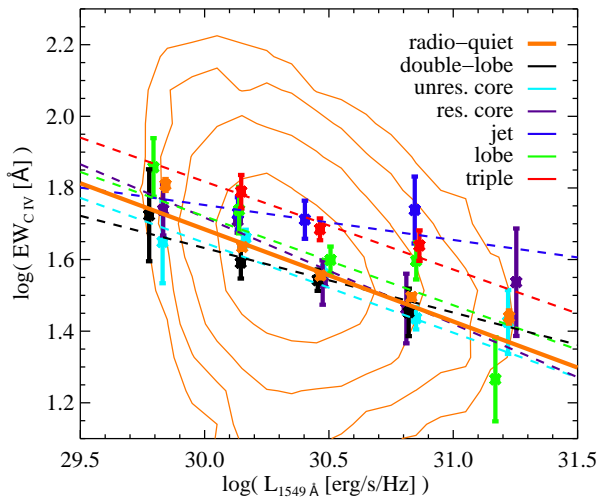}
\figcaption{\label{fig:baldwin}
The Baldwin effect across the quasar radio morphology classes: \ewciv\ as a function of continuum luminosity at 1549 \AA.  Symbols correspond to median values in bins with a width of 0.4 in log($L_\nu$) from 29.5 to 31.5.}
\end{figure}


Figure~\ref{fig:linecorr OIII} shows the distribution of \ewoiii\ as a function of $R$ and \ri.  We do not observe any significant correlations with $R$ nor with \ri.  We see a trend in the full radio sample for \ewoiii\ to increase with \ri, but with a significance of only $1.5\sigma$; no individual morphology classes show any stronger trend.
  
The weak positive correlation between narrow line EW and core-boosting parameter conflicts with the results of \citet{baker97} and \citet{jacksonNature89}, who observed an anti-correlation of \ewoiii\ with $R$.  We note that the quasar samples used in those studies were an order of magnitude smaller than the sample presented in this paper.  As for Mg\,II reported above, our sample matches the Baker sample in redshift, but not in radio brightness (our sources are much fainter).  While the orientation-based theories discussed in the beginning of Section~\ref{sec:line_profiles} suggest that \ewoiii\ is anti-correlated with core-boosting, we see no evidence of such a trend in our data.

\subsection{The \civline\ Baldwin Effect}

The Baldwin effect is a well-known anti-correlation between broad-line width and optical continuum luminosity.  It was first observed in C\,IV \citep{baldwinEffect}, and has since been discovered in many other lines \citep[e.g.,][and references therein]{osmerShields99,wu09}.  In this section, we compare the Baldwin effect in our sample to previous observations.

Traditionally, the Baldwin effect has been measured as \ewciv\ vs monochromatic luminosity at 1450 \AA.  Here we use the continuum level at the rest-wavelength of the C\,IV line, 1549\AA\footnote{Continuum measurements at 1549\AA\ are from the SDSS.}.  This modification should not significantly affect the slope of our Baldwin effect measurement because the luminosities at these two wavelengths are strongly correlated \citep{yeeOke78,shuder81}.

Figure~\ref{fig:baldwin} shows the Baldwin effect we observe in each quasar morphology class.  Previously observed slopes are typically around $-0.23$ \citep[e.g.]{osmer94,green96}.  Our radio-quiet sample has a slightly steeper slope of $B=-0.257\pm0.005$.  The slopes for the radio quasars are consistent with this value, ranging from $B=-0.180\pm0.085$ to $B=-0.297\pm0.068$ for all classes except jet sources.  The jet class, with a slope of $B=-0.098\pm0.114$, has a slope consistent with zero.

It has been noted that the Baldwin effect appears to be shallower for lower-luminosity quasars.  For example, \citet{dietrich02} observed a shallower slope of $-0.14\pm0.02$ in a sample of quasars spanning six orders of magnitude in luminosity, but observed a slope of $-0.20\pm0.03$ when limiting their sample to quasars with $L_\nu(1450$\AA$)\gtrsim5\times10^{28}$~erg~s$^{-1}$~Hz$^{-1}$ \citep[but see also][]{wu09}.  For most of our sources, our slope values are consistent with the higher-luminosity slope of \citet{dietrich02}; all of our quasars are brighter than the luminosity limit they used to separate their sample into high- and low-luminosity samples.




\section{Summary and Conclusions}
\label{sec:discussion}

We have compiled the largest sample to date of radio quasars with high-quality spectra and reliable visual classification by radio morphology.  The sample consists of 4714 radio quasars from the SDSS DR5 spectroscopic survey with a 20~cm FIRST detection brighter than 2~mJy within 2\arcsec\ of the optical position.  We have used the sample to investigate how optical spectral properties of a quasar sample depend on a purely radio-based selection.  One investigative technique involves spectral composites generated for individual radio morphology classes.  We also compared two measures of quasar core-boosting, $R$ and \ri, and searched for correlations between these parameters and spectral line equivalent widths.  Data products available in the electronic version of the paper are listed in the Appendix.

Our main results are as follows:

\begin{enumerate}

\item We found that generally it is not true that radio quasars have redder optical colors than radio-quiet quasars.  Among a $z\lesssim2.7$ sample, radio quasars with unresolved core morphology contain a higher fraction of extremely-reddened objects than do radio quasars with resolved core, triple, lobe, jet, or double-lobe morphology, as well as radio-quiet quasars.  These results also suggest that the SDSS color-selection criteria for targeting quasars are not strongly ($<0.1$~mag) biased, because all other radio quasar classes have observed color distributions similar to that of radio-quiet quasars.

\item We found that \ri\ and $R$, two parameters that have been proposed as measures of quasar radio-core boosting, are correlated.  This observation indicates that some physical factor drives both parameters; core-boosting models suggests that orientation is the driving factor.  The scatter in the observed correlation indicates that another effect is also involved; external environment or age are two such possible factors.

\item In contrast to \citet{baker97}, we did not find significant correlations between $R$ and emission line EW for C\,IV $\lambda1549$, Mg\,II $\lambda2799$, or [O\,III] $\lambda\lambda4959/5007$.  These results suggest that orientation does not govern EW of quasar emission lines, or that the line emission and the optical continuum have the same dependence on orientation, as might be the case if both the lines and continuum originate in a disk-like geometry.  On the other hand, we did observe positive correlations between \ri\ and broad line EW for the full radio quasar sample and also for some of the individual morphology classes, including core- and triple-morphology sources.  If \ri\ is taken as a statistical indicator of orientation, then these observations suggest that orientation {\it does} play a role in determining broad line EW in radio quasars.  

At first glance, the above conclusions appear to be inconsistent, given that both $R$ and \ri\ should correlate with line-of-sight orientation.  However, \citet{willsBrotherton95} and \citet{kharbEtal10} suggested that \ri\ is a stronger orientation indicator than $R$, because the latter parameter also depends significantly on environment and age, as discussed in Section~\ref{sec:core-boosting}.  Our results are consistent with the idea that line EW depends on orientation, because of the observed correlations with \ri.  A corollary of that conclusion is a possible confirmation that \ri\ depends more strongly on orientation than does $R$, because of the lack of observed correlations between EW and $R$.  In that case, the observed correlations may be the result of an anisotropically-emitting BLR  \citep[e.g.,][]{horneEtal03}.  However, our results do not constitute definitive proof of these conclusions.  It is also possible that a physical parameter other than orientation is the driving factor of the \ri--EW correlations.


\renewcommand{\arraystretch}{1.6}
\setlength{\tabcolsep}{3pt}

\begin{deluxetable*}{cclcrrrr}
\tablewidth{6in}
\tablecaption{\label{table:fits}Linear regression results in Section~\ref{sec:line_profiles}}
\tablehead{\colhead{$Y$\tablenotemark{$\dagger$}} & \colhead{$X$\tablenotemark{$\dagger$}} &
  \colhead{class} & \colhead{\# of sources} & \colhead{$A$\tablenotemark{$\dagger$}} &
  \colhead{$B$\tablenotemark{$\dagger$}} & \colhead{Pearson\tablenotemark{$\ddagger$}} & \colhead{Figure}}
\startdata
\enddata
EW Mg\,II & $R$ & lobe & 261 & $1.49_{-0.043}^{+0.043}$ & $0.029_{-0.022}^{+0.023}$ & $0.086_{-0.066}^{+0.068}$ & \ref{fig:linecorr MgII} \\
 & & triple & 420 & $1.61_{-0.011}^{+0.011}$ & $-0.004_{-0.015}^{+0.016}$ & $-0.012_{-0.050}^{+0.052}$ & \\
\hline 
 & & {\bf all radio} & {\bf 2929} & ${\bf 1.42_{-0.011}^{+0.012}}$ & ${\bf 0.053_{-0.006}^{+0.005}}$ & ${\bf 0.179_{-0.018}^{+0.018}}$ & \\
 & & {\bf unresolved core} & {\bf 1602} & ${\bf 1.42_{-0.016}^{+0.015}}$ & ${\bf 0.034_{-0.008}^{+0.007}}$ & ${\bf 0.112_{-0.025}^{+0.024}}$ & \\
EW Mg\,II & \ri & {\bf resolved core} & {\bf 384} & ${\bf 1.38_{-0.024}^{+0.024}}$ & ${\bf 0.070_{-0.012}^{+0.012}}$ & ${\bf 0.309_{-0.050}^{+0.046}}$ & \ref{fig:linecorr MgII} \\
 & & {\bf jet} & {\bf 126} & ${\bf 1.33_{-0.075}^{+0.073}}$ & ${\bf 0.103_{-0.028}^{+0.029}}$ & ${\bf 0.332_{-0.090}^{+0.087}}$ & \\
 & & lobe & 265 & $1.48_{-0.039}^{+0.040}$ & $0.032_{-0.020}^{+0.020}$ & $0.096_{-0.060}^{+0.062}$ & \\
 & & {\bf triple} & {\bf 491} & ${\bf 1.41_{-0.030}^{+0.032}}$ & ${\bf 0.100_{-0.015}^{+0.015}}$ & ${\bf 0.294_{-0.044}^{+0.042}}$ & \\
\hline
EW C\,IV & $R$ & lobe & 119 & $1.55_{-0.074}^{+0.073}$ & $0.034_{-0.037}^{+0.037}$ & $0.085_{-0.092}^{+0.092}$ & \ref{fig:linecorr CIV} \\
 & & triple & 109 & $1.70_{-0.023}^{+0.022}$ & $0.028_{-0.032}^{+0.030}$ & $0.090_{-0.106}^{+0.102}$ & \\
\hline
 & & {\bf all radio} & {\bf 1398} & $\bf{1.40_{-0.023}^{+0.023}}$ & $\bf{0.086_{-0.011}^{+0.011}}$ & $\bf{0.215_{-0.026}^{+0.026}}$ & \\
 & & {\bf unresolved core} & {\bf 945} & $\bf{1.37_{-0.030}^{+0.030}}$ & $\bf{0.082_{-0.014}^{+0.014}}$ & $\bf{0.195_{-0.033}^{+0.032}}$ & \\
EW C\,IV & \ri & {\bf resolved core} & {\bf 138} & $\bf{1.40_{-0.054}^{+0.054}}$ & $\bf{0.111_{-0.028}^{+0.026}}$ & $\bf{0.341_{-0.082}^{+0.074}}$ & \ref{fig:linecorr CIV} \\
 & & jet & 46 & $1.63_{-0.116}^{+0.111}$ & $0.034_{-0.041}^{+0.043}$ & $0.139_{-0.169}^{+0.170}$ & \\
 & & lobe & 119 & $1.56_{-0.079}^{+0.073}$ & $0.033_{-0.036}^{+0.040}$ & $0.082_{-0.090}^{+0.100}$ & \\
 & & {\bf triple} & {\bf 130} & $\bf{1.48_{-0.067}^{+0.067}}$ & $\bf{0.110_{-0.032}^{+0.032}}$ & $\bf{0.298_{-0.084}^{+0.084}}$ & \\
\hline
EW [O\,III] & $R$ & lobe & 60 & $1.38_{-0.156}^{+0.145}$ & $0.002_{-0.083}^{+0.083}$ & $0.003_{-0.138}^{+0.136}$ & \ref{fig:linecorr OIII} \\
 & & triple & 137 & $1.57_{-0.042}^{+0.041}$ & $-0.083_{-0.059}^{+0.062}$ & $-0.116_{-0.084}^{+0.087}$ & \\
\hline
 & & all radio & 840 & $1.44_{-0.036}^{+0.037}$ & $0.032_{-0.019}^{+0.019}$ & $0.059_{-0.035}^{+0.035}$ & \\
 & & unresolved core & 436 & $1.48_{-0.050}^{+0.049}$ & $-0.003_{-0.027}^{+0.027}$ & $-0.005_{-0.049}^{+0.047}$ &\\
EW [O\,III] & \ri & resolved core & 125 & $1.39_{-0.084}^{+0.081}$ & $0.081_{-0.047}^{+0.046}$ & $0.162_{-0.094}^{+0.090}$ & \ref{fig:linecorr OIII} \\
 & & jet & 32 & $1.32_{-0.338}^{+0.338}$ & $0.063_{-0.132}^{+0.128}$ & $0.102_{-0.215}^{+0.199}$ & \\
 & & lobe & 62 & $1.34_{-0.157}^{+0.159}$ & $0.032_{-0.087}^{+0.088}$ & $0.052_{-0.140}^{+0.138}$ & \\
 & & triple & 161 & $1.45_{-0.103}^{+0.104}$ & $0.078_{-0.051}^{+0.050}$ & $0.127_{-0.083}^{+0.078}$ & \\
\hline
 & & {\bf radio-quiet} & {\bf 21,067} & ${\bf 9.40_{-0.160}^{+0.153}}$ & ${\bf -0.257_{-0.005}^{+0.005}}$ & ${\bf -0.336_{-0.006}^{+0.006}}$ & \\
 & & {\bf all radio} & {\bf 1388} & ${\bf 9.61_{-0.705}^{+0.720}}$ & ${\bf -0.264_{-0.023}^{+0.023}}$ & ${\bf -0.294_{-0.025}^{+0.025}}$ & \\
 & & {\bf unresolved core} & {\bf 937} & ${\bf 9.18_{-0.975}^{+0.901}}$ & ${\bf -0.251_{-0.030}^{+0.032}}$ & ${\bf -0.268_{-0.031}^{+0.033}}$ & \\
 & & {\bf resolved core} & {\bf 138} & ${\bf 10.64_{-2.021}^{+2.103}}$ & ${\bf -0.297_{-0.069}^{+0.067}}$ & ${\bf -0.369_{-0.081}^{+0.080}}$ & \\
EW C\,IV & $L_\nu$(1549\AA) & jet & 45 & $4.68_{-3.417}^{+3.449}$ & $-0.098_{-0.114}^{+0.113}$ & $-0.146_{-0.167}^{+0.168}$ & \ref{fig:baldwin} \\
 & & {\bf lobe} & {\bf 118} & ${\bf 9.16_{-1.984}^{+2.055}}$ & ${\bf -0.248_{-0.067}^{+0.066}}$ & ${\bf -0.342_{-0.087}^{+0.091}}$ & \\
 & & {\bf triple} & {\bf 130} & ${\bf 9.19_{-2.025}^{+2.076}}$ & ${\bf -0.246_{-0.068}^{+0.067}}$ & ${\bf -0.327_{-0.089}^{+0.089}}$ & \\
 & & double-lobe & 87 & $7.03_{-2.643}^{+2.566}$ & $-0.180_{-0.084}^{+0.087}$ & $-0.244_{-0.111}^{+0.119}$ & \\
\enddata
\tablecomments{Distributions with significant correlations are shown in bold text.  We define a significant correlation as having slope $B$ and correlation coefficient that are non-zero with at least 3$\sigma$ certainty.  We exclude statistical analysis for samples with fewer than 20 sources.}
\tablenotetext{$\dagger$}{Parameters given in Equation~\ref{eq:fits}.}
\tablenotetext{$\ddagger$}{Value of Pearson correlation coefficient reported by the code of \citet{kelly07}.}
\end{deluxetable*}

\item We observed the Baldwin effect (the anti-correlation between \ewciv\ and continuum luminosity) in both radio and radio-quiet quasars.  We find, for most morphology classes, slopes of approximately $-0.25$.  Owing to large uncertainties, the slope for the double-lobed classes is not strong enough to be considered a significant trend, while the slope for the jet classes is consistent with zero.  For all other morphology classes (the vast majority of the sample), the observed trends are significant; these results are generally consistent with previous observations of the Baldwin effect in high-luminosity quasar samples.


\end{enumerate}

If \ri\ is a statistical indicator of orientation as proposed, our results suggest a change in quasar emission line EW with viewing angle.  For isotropic line emission, the observed correlations imply that the continuum emission increases as the viewing angle to the radio-jet axis increases.  This result is the opposite of what has been observed in earlier, much smaller, samples of quasars: \citet{browneMurphy87}, \citet{bakerHunstead95}, and \citet{baker97} observed an anti-correlation between core-dominance and EW$_\mathrm{MgII}$, while \citet{jacksonNature89} and \citet{baker97} observe a similar anti-correlation among the narrow emission lines.  Basic core-boosting and disk-inclination models predict an increase in continuum emission for small viewing angles (with respect to the radio-jet axis), which would lead to smaller equivalent widths for isotropically emitted lines from quasars with high core-dominance.  The evidence for increased broad line EW with \ri\ in this sample is strong.  It is possible that the interpretation of \ri\ as an orientation indicator is not correct, although its correlation with $R$ argues against this conclusion.  Our results may indicate, therefore, that the broad lines are not emitted isotropically.  A significantly flattened broad line region (BLR) might produce the observed trends with \ri.  In that case, these trends would imply a quite anisotropic distribution of BLR clouds, or perhaps a BLR that is associated with the accretion disk.

\acknowledgements

We extend thanks to Patrick Hall, Brandon Kelly, Rob Gibson, Markos Georganopoulos, and Gordon Richards for useful discussions.  Suggestions from an anonymous referee helped to greatly improve the final version of this paper.  This material is based upon work supported under a National Science Foundation Graduate Research Fellowship, and by NSF grant AST-0507259 to the University of Washington.  D.P.S. acknowledges support from NSF grant AST-0607634.

The National Radio Astronomy Observatory is a facility of the National Science Foundation operated under cooperative agreement by Associated Universities, Inc.

Funding for the SDSS and SDSS-II has been provided by the Alfred P. Sloan Foundation, the Participating Institutions, the National Science Foundation, the U.S. Department of Energy, the National Aeronautics and Space Administration, the Japanese Monbukagakusho, the Max Planck Society, and the Higher Education Funding Council for England. The SDSS Web Site is http://www.sdss.org/. 

The SDSS is managed by the Astrophysical Research Consortium for the Participating Institutions. The Participating Institutions are the American Museum of Natural History, Astrophysical Institute Potsdam, University of Basel, University of Cambridge, Case Western Reserve University, University of Chicago, Drexel University, Fermilab, the Institute for Advanced Study, the Japan Participation Group, Johns Hopkins University, the Joint Institute for Nuclear Astrophysics, the Kavli Institute for Particle Astrophysics and Cosmology, the Korean Scientist Group, the Chinese Academy of Sciences (LAMOST), Los Alamos National Laboratory, the Max-Planck-Institute for Astronomy (MPIA), the Max-Planck-Institute for Astrophysics (MPA), New Mexico State University, Ohio State University, University of Pittsburgh, University of Portsmouth, Princeton University, the United States Naval Observatory, and the University of Washington. 

\appendix

\section{Optically-faint triple-morphology quasars}
\label{sec:optically-faint triples}

We discuss here the construction of the sample of ``optically-faint triple-morphology quasars": triple-morphology sources with no optical detection, or with an optical source too faint for the SDSS spectroscopic quasar sample.  We compile this sample in order to determine whether optically-faint triples have a similar $R$ distribution to optically-bright triples.  The comparison of the two distributions represents one step toward demonstrating that the observed correlation between $R$ and \ri\ is an intrinsic property of radio quasars (\S\ref{sec:core-boosting}).

We drew candidates from the radio catalog of KI08, which identifies all sources from the 20 cm NRAO-VLA Sky Survey (NVSS) \citep{nvss} with three FIRST matches within 120\arcsec.  (Only one triple quasar in our main sample is undetected by NVSS.)  The initial selection criteria, based on parameters in the KI08 catalog, are: {\tt matchflag\_nvss}~=~$-1$, {\tt matchflag\_first}~=~1, {\tt matchflag\_120}~=~3.  These criteria select unique NVSS sources with 3 FIRST matches within 120\arcsec.

\begin{figure}
\plotone{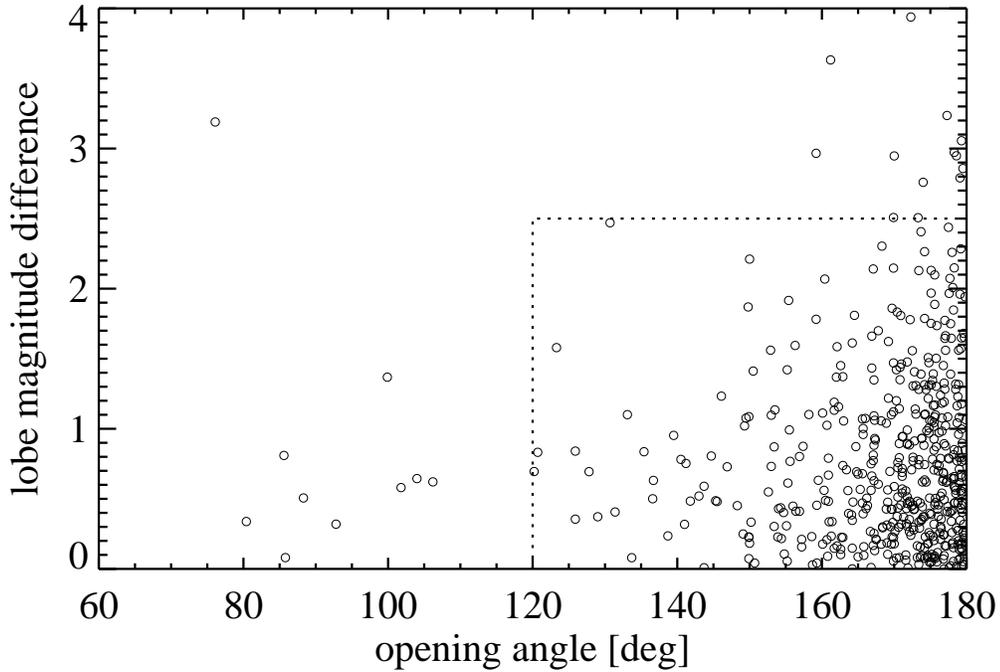}
\figcaption{\label{fig:optically-faint triples}
The distribution of lobe--core--lobe opening angle and lobe--lobe flux density ratio for 519 triple-morphology quasars with well-determined lobe identifications (symbols).  The dotted lines indicate the final selection criteria for this sample: an opening angle greater than 120$\degr$ and a lobe--lobe magnitude difference of less than 2.5, corresponding to a ratio of $<10$ in flux density.}
\end{figure}

Using simple geometry, we eliminated all configurations where the FIRST triplet is arranged in an acute triangle on the sky.  Such configurations are almost certainly spurious: extremely few double-lobed quasars have an opening angle as small as 90$\degr$, as observed by dV06 and confirmed in our main sample triples (Figure~\ref{fig:optically-faint triples}).  We identify the core component as the one ``in-between" the other two FIRST sources.  Following our main sample selection, we require $S_\mathrm{core}>2$~mJy.  To further remove spurious sources, we require that the core source be unresolved in the FIRST survey: $S_\mathrm{20}/S_\mathrm{peak}\lesssim1.23$ (KI08).

We then eliminated configurations with an optical core bright enough for SDSS targeting: we matched the cores to DR6 within 2\arcsec, and removed any with an optical match brighter than $i=19.1$, as well as others which already had an SDSS spectrum.

We decided the final selection criteria by comparing parameters of the remaining 4361 candidates to the main sample of triples.  Figure~\ref{fig:optically-faint triples} shows the lobe---core---lobe opening angle and lobe---lobe magnitude differences (flux density ratios) for the ``optically-faint" and ``optically-bright" triples.  Based on the distribution of the optically-bright sample, we require that the optically-faint candidates have an opening angle of at least 120$\degr$ and a lobe---lobe flux density ratio between 0.1 and 10 (corresponding to a magnitude difference of less than 2.5 mag).  The final sample of optically-faint triples contains 2700 sources.

Of the 2700 optically-faint triples, 719 have an SDSS object located within 2\arcsec\ of the core.  These sources, listed in Table~\ref{table:optically-faint triples}, are excellent candidates for optical quasars, but were too faint for SDSS spectroscopy.

\begin{deluxetable}{rrrccccc}
\tablewidth{0in}
\tablecaption{\label{table:optically-faint triples}
The 719 optically-faint triples in the SDSS}
\tablehead{\multicolumn{2}{c}{SDSS position} & \colhead{FIRST} & \multicolumn{5}{c}{SDSS model magnitudes}\\
\multicolumn{2}{c}{R.A.~~(J2000)~~Dec} & \colhead{flux density [mJy]} & \colhead{$u$} & \colhead{$g$} & \colhead{$r$} & \colhead{$i$} & \colhead{$z$}}
\startdata
   2.020341 &   -0.107590 &     6.32 &  22.73 &  22.71 &  21.97 &  21.27 &  20.62 \\
   3.407161 &    0.653116 &    43.68 &  25.13 &  22.62 &  21.86 &  20.91 &  20.67 \\
  11.999740 &   -8.890384 &     2.23 &  24.24 &  23.59 &  21.37 &  20.11 &  19.82 \\
  25.386895 &    1.011542 &     3.49 &  22.78 &  21.69 &  21.27 &  21.13 &  20.79 \\
  26.709920 &    0.628083 &     2.91 &  22.68 &  22.50 &  20.79 &  19.78 &  18.98 \\
  28.002647 &   -0.487744 &    24.69 &  23.08 &  23.77 &  23.10 &  24.12 &  20.95 \\
  29.117733 &   -9.749173 &    28.06 &  24.67 &  22.03 &  21.36 &  21.02 &  21.34 \\
  29.719878 &    1.025816 &    59.42 &  21.53 &  20.81 &  20.30 &  19.96 &  19.44 \\
  32.471813 &   -0.962489 &     3.49 &  23.68 &  22.44 &  22.23 &  21.36 &  20.97 \\
  33.254807 &   -0.304166 &    52.80 &  22.53 &  22.59 &  21.96 &  21.77 &  21.48 \\
\enddata
\tablecomments{This table is available in full in the electronic version of this paper.}
\end{deluxetable}

\section{Data Products Available for Download}
\label{app:data products}

The following data products are available in the electronic version of this paper.

\begin{enumerate}

\item Table~\ref{table:sample} provides the set of 4714 radio quasars with visually-confirmed SDSS redshifts and radio morphological classifications.  The main radio morphology classes are {\it core}, {\it lobe}, {\it jet}, {\it triple}, {\it knotty}.  Example images are shown in Figure~\ref{fig:mosaic}, and the number of quasars in each class are listed in Table~\ref{table:numbers}.  Triple quasars were sub-classified as {\it core-dominated}, {\it lobe-dominated}, or {\it irregular}; core quasars were sub-classified as {\it resolved} or {\it unresolved}.

\item Table~\ref{table:double-lobed} provides a list of 317 double-lobed quasar candidates.  These sources are quasars with SDSS spectra but with no radio detection within 2\arcsec\ of the optical position.  They are identified by typically symmetric radio emission on either side of the core.  Detailed selection criteria are discussed in Section~\ref{subsec:doubles}.

\item Table~\ref{table:optically-faint triples} lists 719 optically-faint triple-morphology quasars in the SDSS, with selection criteria described in the Appendix.  These sources were identified purely on the basis of their radio emission, and have an SDSS photometric source within 2\arcsec\ of the radio core.  These sources are not spectroscopically-confirmed quasars, as they are too faint for SDSS targeting, but many of them may qualify for spectroscopy in the SDSS~III Baryon Oscillation Spectroscopic Survey \citep{eisenstein11}.

\item Tables~\ref{table:composites_main} and~\ref{table:composites_triple} provide median composite SDSS spectra for the radio-quiet quasars, for the entire radio quasars sample, for individual radio morphology classes (resolved core, unresolved core, triple, jet, lobe), and for triple quasars grouped by rest-frame $R$.

\end{enumerate}

\bibliography{./bibliography}

\end{document}